\documentclass[prb,twocolumn,aps,amsmath,amssymb,amsfonts,superscriptaddress,preprintnumbers,floatfix]{revtex4}
\usepackage[german,american,english]{babel}
\usepackage{graphicx}
\usepackage{graphics}
\usepackage{dcolumn}
\usepackage{bm}
\usepackage{latexsym}
\usepackage{amssymb}
\usepackage{amsmath}
\usepackage{amsfonts}
\usepackage{layout}
\usepackage{verbatim}
\usepackage{epsfig}

\newcommand {\ket} [1] {| #1 \rangle}

\newcommand {\dbkt} [2] {\langle #1 | #2 \rangle}
\newcommand {\tbkt} [3] {\langle #1 | #2 | #3 \rangle}

 \newcommand {\beq}{\begin{equation}}
\newcommand {\eeq}{\end{equation}}
\begin{document}
\title{Quantum dot spin qubits in Silicon: Multivalley physics}
\author{Dimitrie Culcer}
\affiliation{Condensed Matter Theory Center, Department of Physics, University of Maryland, College Park MD 20742-4111}
\author{{\L}ukasz Cywi{\'n}ski}
\affiliation{Condensed Matter Theory Center, Department of Physics, University of Maryland, College Park MD 20742-4111}
\affiliation{Institute of Physics, Polish Academy of Sciences, al.~Lotnik{\'o}w 32/46, PL 02-668 Warszawa, Poland}
\author{Qiuzi Li}
\affiliation{Condensed Matter Theory Center, Department of Physics, University of Maryland, College Park MD 20742-4111}
\author{Xuedong Hu}
\affiliation{Condensed Matter Theory Center, Department of Physics, University of Maryland, College Park MD 20742-4111}
\affiliation{Joint Quantum Institute, Department of Physics, University of Maryland, College Park MD 20742-4111}
\affiliation{Department of Physics, University at Buffalo, SUNY, Buffalo, NY 14260-1500}
\author{S.~Das Sarma}
\affiliation{Condensed Matter Theory Center, Department of Physics, University of Maryland, College Park MD 20742-4111}
\begin{abstract}
Research on Si quantum dot spin qubits is motivated by the long spin coherence times measured in Si, yet the orbital spectrum of Si dots is
increased as a result of the valley degree of freedom. The valley degeneracy may be lifted by the interface potential, which gives rise to a
valley-orbit coupling, but the latter depends on the detailed structure of the interface and is generally unknown {\it a priori}. These facts
motivate us to provide an extensive study of spin qubits in Si double quantum dots, accounting fully for the valley degree of freedom and
assuming no prior knowledge of the valley-orbit coupling. For single-spin qubits we analyze the spectrum of a multivalley double quantum dot,
discuss the initialization of one qubit, identify the conditions for the lowest energy two-electron states to be a singlet and a triplet well
separated from other states, and determine analytical expressions for the exchange splitting. For singlet-triplet qubits we analyze the
single-dot spectrum and initialization process, the double-dot spectrum, singlet-triplet mixing in an inhomogeneous magnetic field and the
peculiarities of spin blockade in multivalley qubits. We review briefly the hyperfine interaction in Si and discuss its role in spin blockade in
natural Si, including intravalley and intervalley effects. We study the evolution of the double-dot spectrum as a function of magnetic field. We
address briefly the situation in which the valley-orbit coupling is different in each dot due to interface roughness. We propose a new
experiment for measuring the valley splitting in a \textit{single} quantum dot. We discuss the possibility of devising other types of qubits in
Si QDs, and the size of the intervaley coupling due to the Coulomb interaction.
\end{abstract}
\maketitle

\section{Introduction}

A practical QC architecture requires scalability and long coherence times. Spin-based solid state quantum bits (qubits) are known to have long
coherence times,\cite{Feher_PR59, Petta_Science05, Koppens_PRL08} while also offering the promise of scalability, and are natural building
blocks for quantum computation. The conventional criterion establishing an acceptable coherence time for fault-tolerant quantum computing is for
it to allow $10^4$ operations. With the operation time set by the gating time $\approx 1$ns the coherence time required is $\approx 10\mu$s.

Phosphorus donor nuclei in Silicon have been known since the 1950s to have some of the best spin coherence properties in
solids,\cite{Feher_PR59, Tyryshkin_PRB03, Tyryshkin_JPC06, Hanson_RMP07} and were therefore proposed as viable candidates for
qubits.\cite{Kane_Nature98} However, fabrication of ordered and gated donor arrays and coherent control over donor electrons has turned out to
be extremely difficult (note nevertheless outstanding recent developments. \cite{Mottonen_PRB10, Lim_APL09, Lim_SingleElectron_APL09})

Electron spin qubits in quantum dots (QDs), \cite{Loss_PRA98} particularly in GaAs, have been actively studied in the past decade, and coherence
times in excess of $ \approx100\mu$s have been achieved.\cite{Bluhm_LongCoh_10, Petta_Science05, Koppens_PRL08} Various schemes exist for
implementing spin qubits in QDs. For the original single-spin qubits, one electron spin is initialized in each dot in a QD
array,\cite{Loss_PRA98} with spin up and down representing the two states of a qubit. One-qubit rotations are accomplished using magnetic
resonance techniques and two-qubit rotations by means of a pulsed exchange interaction. One can also use selected two-spin states to encode a
logical qubit: in singlet-triplet qubits it is the singlet and unpolarized triplet states that are used for encoding.  Here exchange coupling
gives the splitting of the two-spin states making up the qubit.\cite{Levy_PRL02, Petta_Science05, Taylor_NP05} One-qubit rotations are currently
carried out using an inhomogeneous magnetic field across the two dots and pulsed exchange splitting. Two-qubit operations use the different
dipole moments of the singlet and triplet states. \cite{Taylor_NP05} Larger encoding schemes such as using three-spin states have  also been
proposed,\cite{DiVincenzo_Nature00} aiming at an all-electrically-controlled architecture.

Recent years have seen significant experimental progress involving single-spin properties such as coherent control, coherence, and
measurement.\cite{Ono_Science02, Elzerman_Nature04, Johnson_Nature05, Koppens_Science05, Hanson_RMP07, MarcusGroup_NatureNano07, Shaji_NP08,
Koppens_PRL08, Reilly_PRL08, Barthel_PRL09} In GaAs double quantum dots (DQDs) spin blockade \cite{Ono_Science02} and charge sensors (quantum
point contacts and radio-frequency quantum point contacts)\cite{Field_PRL93} enable observation of single/two-spin
dynamics.\cite{Petta_Science05, Koppens_Science05, Koppens_PRL08, Taylor_PRB07, Hanson_RMP07} Spin coherence in GaAs QDs is mostly limited by
hyperfine interaction with the nuclear spins,\cite{Merkulov_PRB02, Khaetskii_PRB03, Coish_PRB05, Yao_PRB06, Witzel_PRB06, Saikin_PRB07,
Coish_PRB08, Cywinski_PRL09,Cywinski_PRB09} though proposals exist for manipulating nuclear spin states to extend electron spin coherence
times,\cite{Klauser_PRB06, Danon_PRL08, Ramon_PRB07,Stopa_PRB10} and recent experiments\cite{Bluhm_LongCoh_10, Petta_Science05} have shown that
dynamical decoupling schemes \cite{Yao_PRB06, Yao_PRL07, Witzel_PRL07, Uhrig_PRL07, Lee_PRL08, Cywinski_PRB09} could help unwind the nuclear
spin dynamics. Progress has also been made on controlling spins in GaAs QDs by optical means, such as the preservation of spin
coherence,\cite{Greilich_Science06} control of the nuclear spin polarization, \cite{Greilich_Science07, Vink_NP09, Latta_NP09} and ultrafast
spin rotations.\cite{Greilich_NP09} Yet experiment is far from achieving reliable control over the nuclear field in GaAs.

Silicon has outstanding spin coherence properties \cite{Feher_PR59} due to small spin-orbit coupling,\cite{Tahan_PRB05} small hyperfine
interaction with nuclear spins\cite{Assali_preprint10} (which can be reduced by isotopic purification \cite{Witzel_AHF_PRB07}) and absence of
piezoelectric coupling.\cite{Prada_PRB08} The spin coherence time $T_2$ (after Hahn spin echo) for donor electron spins in bulk Si:P has been
reported to range from 60 ms \cite{Tyryshkin_JPC06} to 300 ms. \footnote{S.~Lyon, private communication} This is the longest coherence time
measured in electron spin qubits, and greatly exceeds
the values reported in GaAs QDs, which, after Hahn echo, range from $\sim 1 \mu$s \cite{Petta_Science05, Koppens_PRL08} to $\sim 30
\mu$s.\cite{Bluhm_LongCoh_10} Isotopic purification of Si, where $^{29}$Si nuclei are preferentially removed in favor of $^{28}$Si and $^{30}$Si
nuclei from natural Si, allows enhancement of Si electron spin coherence times.\cite{Abe_PRB04, Tyryshkin_JPC06} The maturity and continuous
innovation in Si microfabrication could be of great help in scaling up a Si-based QC architecture. Work is under way on Si QC architectures
based on donor electron or nuclear spins in a Si:P system,\cite{Kane_Nature98, Vrijen_PRA00} single electron spins in gate-defined quantum dots
in Si/SiGe\cite{Friesen_PRB04, Hayes_09} or Si/SiO$_2$,\cite{Liu_PRB08, Nordberg_MOS_PRB09, Ferrus_09} and holes in SiGe
nanowires.\cite{MarcusGroup_NatureNano07} Experimental progress has been made in both donor-based devices\cite{Andresen_NanoLett07,
Mitic_Nanotech08, Kuljanishvili_NP08, Lansbergen_NP08, Fuhrer_NanoLett09, Stegner_NP06, Angus_NL07,Mottonen_PRB10, Lim_APL09,
Lim_SingleElectron_APL09}, and gate-defined QDs. \cite{Rokhinson_PRL01, Goswami_NP07, Shaji_NP08, Liu_PRB08, Nordberg_MOS_PRB09, Lim_APL09,
Lim_SingleElectron_APL09, Nordberg_APL09}

The valley degree of freedom presents a potential obstacle to spin QC in Si. In bulk Si there are six degenerate conduction band minima near the
X points in the first Brillouin zone. Impurity scattering breaks the symmetry of the lattice and allows coupling between the valleys, and in
samples in which two or more donor centers are present interference becomes possible and is sensitively dependent on the exact atomic locations
of the two donors. \cite{Koiller_PRL01, Wellard_PRB05, Koiller_PRB06} Confinement in the $\hat{\bm z}$-direction and/or uniaxial tensile strain
at the interface cause the valleys perpendicular to the interface to have a smaller energy by several tens of meV than the valleys in the plane
of the interface. A sharp interface further couples the two $z$-valleys by producing a valley-orbit coupling $\Delta$ (a complex number).
\cite{Ando_PRB79} $\Delta$ is generally not known {\it a priori}, is sample-dependent, \cite{Friesen_PRB10} and is different for different
interfaces, such as Si/SiGe and Si/SiO$_2$. Various measurements of $\Delta$ have been reported, \cite{Ando_RMP82, Lai_PRB06, Takashina_PRL06,
Goswami_NP07} and efforts have been devoted to calculating $\Delta$. \cite{Saraiva_PRB09, Saraiva_LargeVOC_10, Friesen_PRB10, Boykin_APL04,
Friesen_PRB04, Nestoklon_PRB06, Srinivasan_APL08} Yet the fact that $\Delta$ is not known and no standard experiment exists to measure it in a
QD at low field is a strong motivating factor for our work.

\begin{figure}[tbp]
\begin{tabular}{lr}
(a) \\
\includegraphics[width=3.0in]{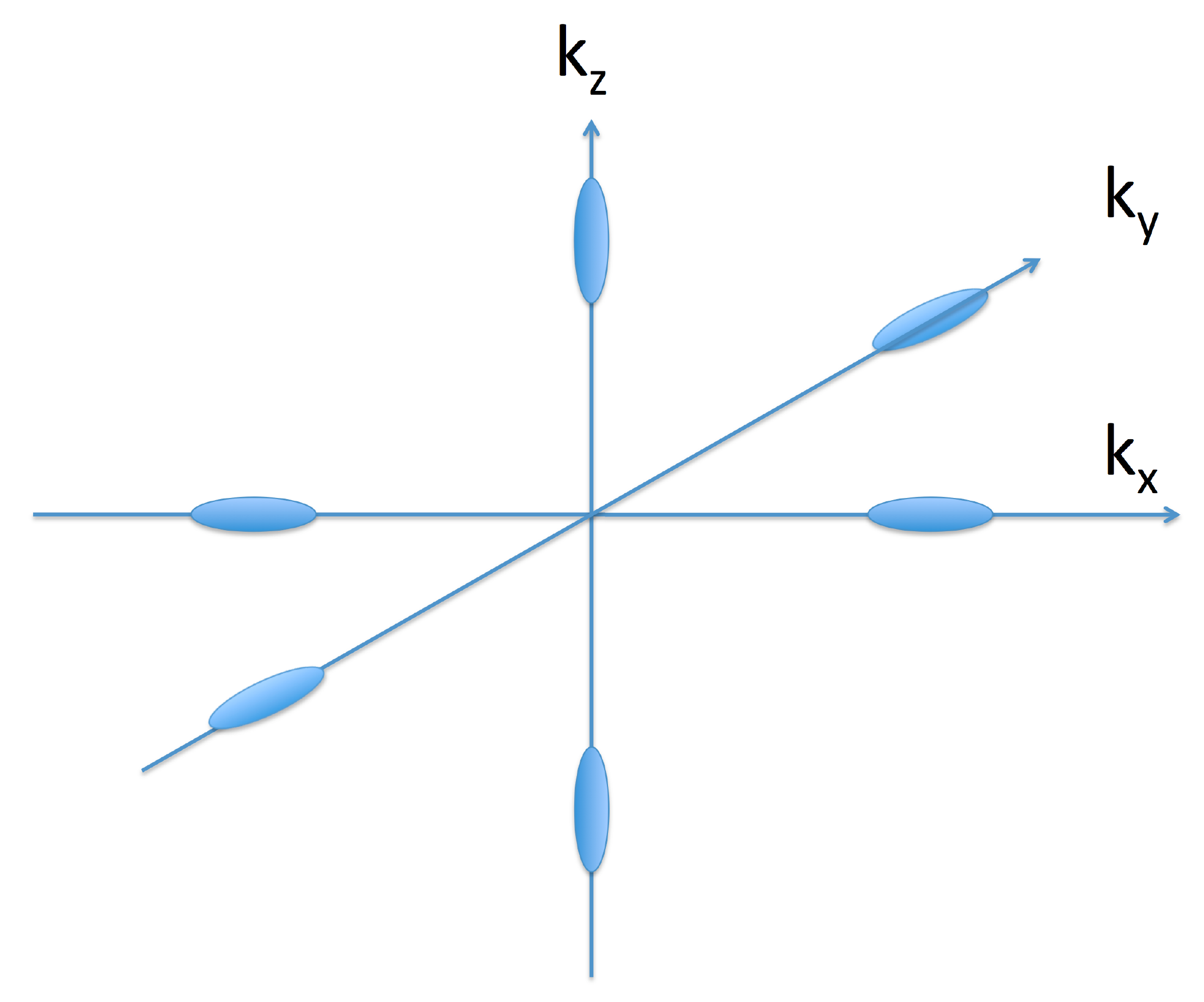}
\end{tabular}
\begin{tabular}{lr}
(b) \\
\includegraphics[width=3.0in]{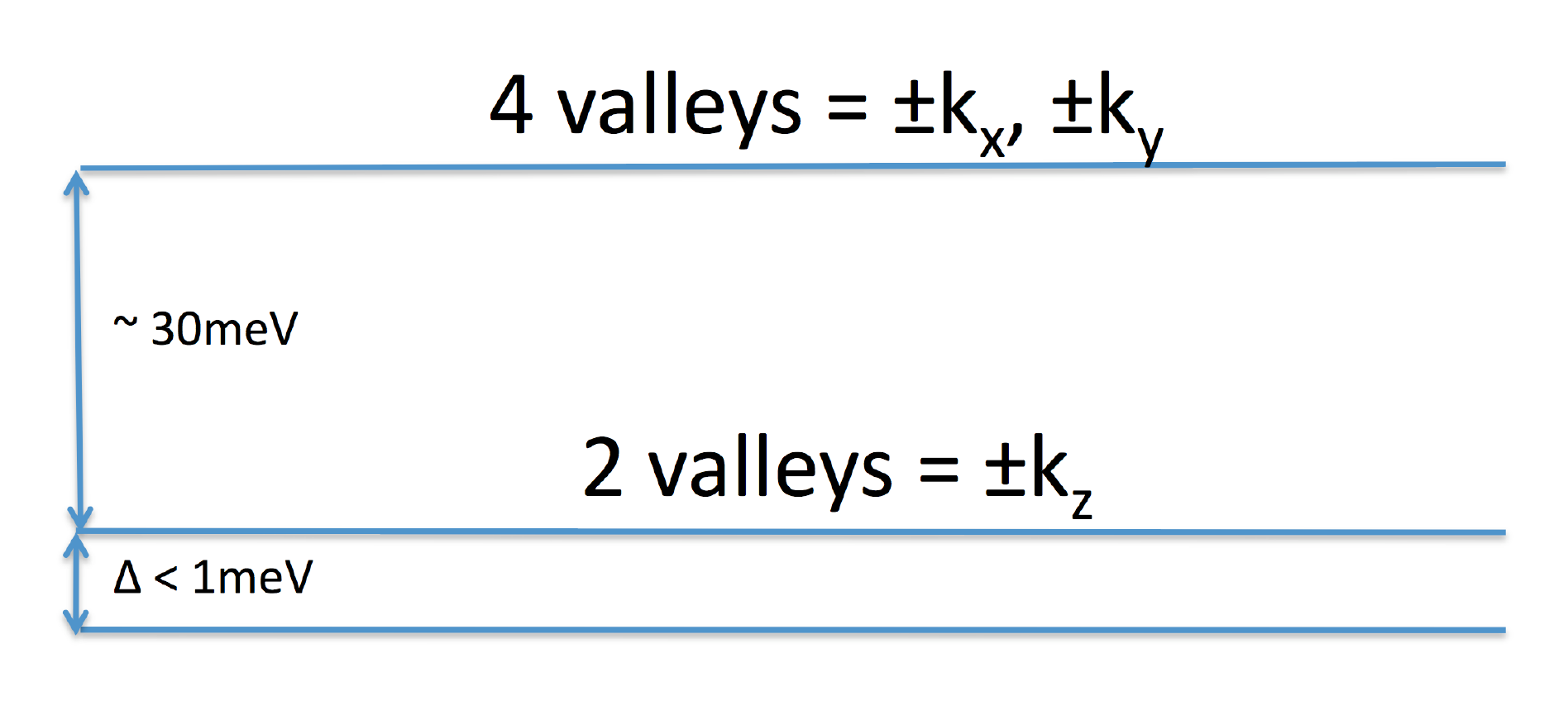}
\end{tabular}
\caption{(color online) Conduction-band valleys in silicon in the bulk and in 2D. (a) In bulk Si there are six equivalent conduction-band minima
positioned along the Cartesian axes. (b) Uniaxial strain at the interface causes the valleys parallel to the interface (i.e. in the $xy$-plane
here) to be higher by several tens of meV than the valleys perpendicular to the interface. The interface potential further splits the
lower-energy valleys by an amount $\Delta$, which present experimental work is striving to determine unambiguously. Based on measurements to
date we expect $\Delta < $1meV.} \label{Si_Valleys}
\end{figure}

In a recent article \cite{Culcer_PRB09} we discussed singlet-triplet qubits in Si QDs, emphasizing the role of the valley degree of freedom.
The focus of Ref.\ \onlinecite{Culcer_PRB09} was on determining the feasibility of an experiment analogous to Ref.~\onlinecite{Petta_Science05}
in a Si/SiO$_2$ (or Si/SiGe \cite{Shaji_NP08}) DQD. It was found that the valley degree of freedom affects qubit initialization, operation and
spin blockade. For $|\Delta| \gg k_BT$ a singlet-triplet qubit can be implemented in the same way as in GaAs. For $|\Delta| \approx k_BT$
initialization will be problematic, since, unlike in GaAs, the energy scale is no longer set by the confinement energy but by the valley
splitting (we note a similar discussion of spin blockade in carbon nanotube quantum dots.\cite{Palyi_PRB09, Palyi_10})

In this work we present a detailed analysis of both single-spin and singlet-triplet qubits in multivalley Si QDs within the effective mass
approximation.  The discussion of valley physics is general and applies to all types of Si QDs, yet when concrete figures are needed we focus on
Si/SiO$_2$. Many aspects of the following considerations apply also to carbon-based QDs, where the valley degree of freedom is present as well.
\cite{Churchill_PRL09,Palyi_PRB09, Palyi_10} We consider first single-spin qubits, i.e. the Loss-DiVincenzo (LDV) architecture.\cite{Loss_PRA98}
We determine the spectrum and the conditions required for implementing single-spin qubits and coupling adjacent qubits by means of the exchange
interaction. We subsequently study singlet-triplet qubits.\cite{Petta_Science05,Taylor_NP05, Culcer_PRB09} Part of the discussion is intended as
a follow up to Ref.\ \onlinecite{Culcer_PRB09}, providing further theoretical details of singlet-triplet qubits including all the relevant
analytical formulas and matrix elements. The remainder contains an extensive discussion of the role of the hyperfine interaction in spin qubits
in natural Si, including intervalley effects, an exhaustive discussion of single-qubit gates and scalability, as well as new experimental
initiatives. We conclude that in both schemes, for any $\Delta$, \textit{single}-qubit operations are feasible. If $|\Delta| \le k_BT$ in
single-spin qubits exchange-based entanglement will not work since exchange between pairs of states with different valley composition is vastly
different. For this reason, the proposed Coulomb interaction-based two-qubit operations in the singlet-triplet scheme will not work either.

The outline of this article is as follows. In Sec.~\ref{sec:DQD} we introduce the DQD Hamiltonian.  In Sec.~\ref{sec:Loss_PRA98} we study the
effect of the valley degree of freedom on single-spin qubits and in Sec.~\ref{sec:ST} we investigate singlet-triplet qubits and the role of the
valley degree of freedom in initialization, coherent manipulations, and readout. We give a brief review of hyperfine interactions in Si, then
discuss intravalley and intervalley effects relevant to singlet-triplet qubits. Interface roughness is discussed in Sec.\ \ref{sec:position},
other types of qubits in Sec.\ \ref{sec:other}, and in Sec.~\ref{sec:identyfying} an experiment is discussed that can estimate the valley
splitting. We also discuss briefly the effect on spin qubits of interface roughness, which may lead to a different valley-orbit coupling in the
two dots. We end with a summary and conclusions.

\section{Model of a double quantum dot}
\label{sec:DQD}

We focus on a DQD in a two-dimensional electron gas (2DEG) in the inversion layer of a MOSFET grown along the $\hat{\bm z}$-direction (which is
also the crystalline (001) direction). The dots are located at ${\bm R}_D = (X_D, 0, 0)$, where $D = L, R$, and $X_R = X_0$ while $X_L = - X_0$
(since the $\hat{\bm z}$ and $-\hat{\bm z}$ valleys are isotropic in the $xy$ plane, we can choose the $\hat{\bm x}$ and $\hat{\bm y}$ axes by
convenience). We do not consider spin-orbit interactions. The DQD Hamiltonian $H = H_0 + \sum_{i=1,2} H_v^{(i)}$, where $H_{v}$ represents the
valley-orbit coupling discussed below and $H_0 = \sum_{i=1,2}\big(T^{(i)} + V_Q^{(i)}\big) + V_{ee}$, where $T$ is the kinetic energy operator
and $V_Q$ the confinement potential
\begin{equation}
V_Q = \displaystyle \frac{\hbar \omega_0}{2a^2} \, \bigg\{\mathrm{Min} [(x-X_0)^2, (x+X_0)^2] + y^2\bigg\} + \frac{\hbar
\omega_z}{2b^2} \, z^2 - eEx,
\end{equation}
with $a$ and $b$ the Fock-Darwin radii for in- and out-of-plane confinement respectively. The external electric field ${\bm E} \parallel
\hat{\bm x}$ is used to bias the DQD and is a crucial tuning knob for singlet-triplet qubits. The Coulomb potential between electrons at ${\bm
r}_1$ and ${\bm r}_2$ is $V_{ee} = e^2/(\epsilon|{\bm r}_1 - {\bm r}_2|)$, where the effective dielectric constant $\epsilon = (\epsilon_{Si} +
\epsilon_{SiO_2})/2$ accounts for the extra screening due to the image charge in SiO$_2$ (for Si/SiGe such averaging is not needed). We use a
quadratic potential to model the growth-direction confinement, which is generally triangular. The in-plane confinement potential for each dot is
harmonic near the center, $V_D (x) = (\hbar \omega_0/2a^2) \, (x - X_D)^2$, with Gaussian ground states.

The confinement along the growth direction splits the six-fold degenerate Si conduction band into a doubly-degenerate branch of lower energy and
a fourfold-degenerate branch of higher energy. The two lower energy valleys $\xi = z, \bar{z}$ are along the direction perpendicular to the
interface at ${\bm k}_\xi = \pm k_0 \, \hat{\bm z}$, where $k_0 = 0.85 (2\pi/a_{Si})$, and the lattice constant $a_{Si} \! = 5.43 {\rm\AA}$. The
ground-state wave functions $D_{\xi}$ satisfy $(T + V_D) \, D_\xi = \varepsilon_0 \, D_\xi$, with
\begin{equation}
D_\xi = F_D ({\bm r} - {\bm R}_D) e^{i{\bm k}_\xi \cdot {\bm r}} u_\xi ({\bm r}),
\end{equation}
where $D = L$ or $R$. The envelope functions are
\begin{equation}
F_D({\bm r} - {\bm R}_D) = \frac{1}{\pi^{3/4} (a^2b)^{1/2}} \, e^{-\frac{(x - X_D)^2}{2a^2}} e^{-\frac{y^2}{2a^2}}e^{-\frac{z^2}{2b^2}},
\end{equation}
where $a = \sqrt{\frac{\hbar}{m_t\omega_0}}$ is the in-plane Fock-Darwin radius, $b$, the growth-direction confinement length, $m_t = 0.191 m_0$
is the in-plane (transverse) Si effective mass, and $m_0$ the bare electron mass.  The lattice-periodic function $ u_\xi ({\bm r}) = \sum_{\bm
K} c^\xi_{\bm K} e^{i{\bm K}\cdot{\bm r}}$, with ${\bm K}$ the reciprocal lattice vectors.

Overlaps of the form $\dbkt{L_\xi}{R_\xi}$ are given by $l = e^{-d^2}$, where $d = X_0/a$. Overlaps of the form $\dbkt{D_\xi}{D_{-\xi}}$ are
suppressed by an exponential of the form $e^{-\frac{b^2Q_z^2}{4}} \ll 1$, where $Q_z = \frac{2\pi n_z}{a_{Si}} - 2k_0$ ($n_z$ is an integer).
With $b \approx 3$nm typically,\cite{Ando_RMP82} $b^2Q_z^2/4 \approx 800$ when $n_z = 0$ and $\approx 150$ when $n_z = 1$.  Matrix elements of
potentials smooth in space and interactions between states from different valleys are suppressed by such prefactors: matrix elements of the form
$\tbkt{D_\xi}{H_0}{D_{-\xi}}$ and $\tbkt{D^{(1)}_\xi D^{(2)}_\xi}{H_0}{D^{(1)}_{-\xi}D^{(2)}_\xi}$. The one exception is the valley-exchange
Coulomb integral $\tilde{j}_v$ below. Two-particle matrix elements in which each electron belongs to a different valley, for example
$\tbkt{D^{(1)}_\xi D^{(2)}_{-\xi}}{H_0}{D^{(1)}_{\xi}D^{(2)}_{-\xi}}$, are of the same order of magnitude as those in which all functions belong
to the same valley, such as $\tbkt{D^{(1)}_\xi D^{(2)}_\xi}{H_0}{D^{(1)}_\xi D^{(2)}_\xi}$. Matrix elements of interactions on the atomic length
scale are not exponentially suppressed.  One prominent example is the hyperfine (hf) interaction between a conduction electron and a $^{29}$Si
nuclear spin (in a natural Silicon substrate about 4.7\% of the nuclei are $^{29}$Si).

Based on multiple experiments conducted over the past 50 years and on the recent calculations (see Sec.~\ref{sec:hfSi} for details), it is
assumed that the hf interaction in Si is dominantly of the contact character, i.e.~the Hamiltonian is given by
\beq
 H_{HF} =  \mathcal{A}_{0}\nu_{0}\sum_i  {\bf S} \cdot {\bf I}_i \delta({\bf r} - {\bf R}_i) \,\, ,
\eeq
where the sum runs over the positions of the $^{29}$Si nuclei, $\mathcal{A}_{0} = \frac{2}{3}\mu_{o}\hbar^{2}\gamma_{S}\gamma_{J}/\nu_{0}$ (with
$\gamma_{S}$ and $\gamma_{J}$ being the electron and nuclear gyromagnetic factor, respectively), $\nu_{0}$ is the unit cell volume and
$\mathbf{I}_{i}$ are the nuclear spins.  The valley-diagonal matrix element is
\beq \tbkt{D_\xi}{H_{HF}}{D_{\xi}} = \sum_{i} \mathcal{A}\nu_{0}|F_{D}({\bf R}_i)|^2 {\bf S} \cdot {\bf I}_i
\eeq
where the hf interaction energy is $\mathcal{A} \! = \! \mathcal{A}_{0} \eta$ with $\eta \! \equiv \! |u_{\xi}({\bf R}_i)|^2 \! \approx \! 160$,
see Ref.~\onlinecite{Assali_preprint10} and references therein ($\mathcal{A} \! \approx \! 2$ $\mu$eV in Si). This Hamiltonian provides a
valley-orbit coupling mechanism, i.e.~we have $\tbkt{D_\xi}{H_{HF}}{D_{-\xi}} = \mathcal{A}_{0}\nu_{0} \sum_{i} D_\xi^*({\bf R}_i) D_{-\xi}({\bf
R}_i) {\bf S} \cdot {\bf I}_i$. The consequences of this coupling are explored below.

The interface potential couples the $z$ and $\bar{z}$ valleys if it is sufficiently sharp on the atomic scale.  This is contained in $H_v$,
which represents the valley-orbit coupling $\Delta$ studied in detail recently.\cite{Saraiva_PRB09, Friesen_PRB10} The magnitude and phase of
$\Delta$ are of great interest in Si research. It has been shown that $|\Delta|$ can range from zero to 0.25 meV (maybe even
larger),\cite{Saraiva_PRB09, Saraiva_LargeVOC_10} and both its magnitude and phase can be controlled by an electric field. In the present work
$H_v$ is a single-particle phenomenological coupling between the valleys $\tbkt{D_\xi}{H_v}{D_{-\xi}} = \Delta \equiv |\Delta|\, e^{-i\phi} $,
with $|\Delta| \!  >\! 0$. Throughout this work $\hbar \omega_z \gg \hbar \omega_0 \gg |\Delta|$.

We diagonalize the single-particle Hamiltonians including $H_v$ and obtain the \textit{valley eigenstates} $D_\pm = (1/\sqrt{2}) \, (D_z \pm
e^{i\phi} D_{\bar {z}})$ with eigenvalues $\varepsilon_0 \pm |\Delta|$ in each of the quantum dots.  We construct orthogonal (Wannier)
single-dot wave-functions \cite{Burkard_PRB99} $\tilde{L}_\xi = \frac{L_\xi - gR_\xi}{\sqrt{1 - 2lg + g^2}}$ and $\tilde{R}_\xi = \frac{R_\xi -
gL_\xi}{\sqrt{1 - 2lg + g^2}}$, where $g = (1 - \sqrt{1 - l^2})/l$, so that $\langle \tilde{R}_\xi | \tilde{L}_\xi \rangle = 0$.  We define
\begin{equation}
\begin{array}{rl}
\displaystyle \tilde{\varepsilon}_0 = & \displaystyle \tbkt{\tilde{D}_\xi}{T + V_D}{\tilde{D}_\xi} \\ [3ex] \displaystyle \tilde{\Delta} = &
\displaystyle \tbkt{\tilde{D}_\xi}{H_v}{\tilde{D}_{-\xi}} \\ [3ex] \displaystyle \tilde{t} = & \displaystyle
\tbkt{\tilde{L}_\xi}{H_0}{\tilde{R}_\xi} + \tbkt{\tilde{L}_\xi \tilde{L}_\xi}{V_{ee}}{\tilde{L}_\xi \tilde{R}_\xi}.
\end{array}
\end{equation}
Orthogonalizing $L_\pm$ and $R_\pm$ we obtain $\tilde{L}_\pm = \frac{L_\pm - gR_\pm}{\sqrt{1 - 2lg + g^2}}$ and $\tilde{R}_\pm = \frac{R_\pm -
gL_\pm}{\sqrt{1 - 2lg + g^2}}$. These are the single-electron states we will use henceforth (e.g. to construct two-electron states.) The orbital
excitation energy is assumed large ($>$ 1meV, or $\gg |\Delta|$.)

\section{Single-spin qubits}
\label{sec:Loss_PRA98}

For single-spin qubits we first consider initialization.  In single-valley systems the lowest energy level in zero magnetic field is
spin-degenerate but orbital-non-degenerate.  Initialization of a definite single-spin state requires a magnetic field to split the spin
degeneracy via the Zeeman interaction, so that selective tunnelling from a reservoir can have high fidelity.\cite{Elzerman_Nature04}
Experimentally in GaAs this process requires magnetic fields of several Tesla due to the inherent smallness of the Bohr magneton, the small
$g$-factor in GaAs ($|g| \sim 0.4$), and the constraint of $\sim$ 100 mK electron temperature in the reservoir. In Si $g \sim 2$, so that at
dilution refrigerator temperatures much smaller magnetic fields will be required than in GaAs, rendering spin initialization a relatively simple
task.

In Si however there are two possibly closely-spaced one-electron levels $\tilde{D}_+$ and $\tilde{D}_-$, separated by $2|\Delta|$.  Initializing
into a definite orbital state is dependent on the size of $|\Delta|$. Since the interaction of spins in states $\tilde{D}_+$ and $\tilde{D}_-$
with a magnetic field is described by identical Zeeman Hamiltonians, single-qubit rotations can be performed no matter which state is
initialized.

\subsection{Two-qubit gates}

In the LDV architecture two-qubit gates are implemented using the exchange interaction between neighboring dots.  Two-electron wave functions
are superpositions of spin singlet and triplet states. Singlet wave function $\ket{\Psi_S} = \ket{\tilde{S}}\ket{\chi_S}$, where
$\ket{\tilde{S}}$ is a two-particle spatially symmetric function and $\ket{\chi_S}$ is the two-spin singlet state given by $\ket{\chi_S} =
(1/\sqrt{2})\, |\uparrow^{(1)} \downarrow^{(2)} - \downarrow^{(1)} \uparrow^{(2)}\rangle$.  The superscript $(i)$ here denotes the $i$-th
electron. Triplet wave functions $\ket{\Psi_T} = \ket{\tilde{T}}\ket{\chi_T}$, where $\ket{\tilde{T}}$ is a two-particle spatially antisymmetric
function, and $\ket{\chi_T}$ represents the spin-triplet states, and is an abbreviation for $\ket{\chi_+} = \ket{\uparrow^{(1)}
\uparrow^{(2)}}$, $\ket{\chi_-} = \ket{\downarrow^{(1)} \downarrow^{(2)}}$, and $\ket{\chi_0} = (1/\sqrt{2})\, |\uparrow^{(1)} \downarrow^{(2)}
+ \downarrow^{(1)} \uparrow^{(2)}\rangle$. The wave functions $\ket{\chi_\pm}$ represent the polarized spin triplets and are given by
$\ket{\uparrow^{(1)} \uparrow^{(2)}}$ and $\ket{\downarrow^{(1)} \downarrow^{(2)}}$ respectively, while $\ket{\chi_0}$ represents the
unpolarized spin triplet and has the form $\ket{\chi_0} = (1/\sqrt{2})\, \ket{\uparrow^{(1)} \downarrow^{(2)} + \downarrow^{(1)}
\uparrow^{(2)}}$.  We will frequently refer to just $\ket{\tilde{S}}$ and $\ket{\tilde{T}}$ as the singlet and triplet states respectively,
because the orbital part is where valleys play an important role.

Translating the LDV architecture to multi-valley systems requires taking into account the valley degree of freedom in the construction of
many-particle wave functions. In general, for a given number $N$ of orbital states one can form $N(N+1)/2$ singlet and $3N(N-1)/2$ triplet
two-electron states.  In a single-valley system the number of orbital states is $N = 1$, resulting in the trivial case of one singlet. In a
single Si QD we have $N = 2$.  For two electrons we have 2(2+1)/2 = 3 singlet states and 3[2(2-1)]/2] = 3 triplet states. In a Si double dot,
when considering the lowest orbital state in each valley, $N=4$, so that there are 4(4+1)/2 = 10 singlet states and 3[4(4-1)]/2 = 18 triplet
states. Therefore there are altogether 28 states for the double dot, of which 10 are spatially symmetric (spin singlets) and 18 are spatially
antisymmetric (spin triplets). The total number of 28 states corresponds to the combination $C^8_2$, where $8 \equiv 2^3$ arises as the product
of a factor of 2 for the spin degree of freedom, a factor of 2 for the valley degree of freedom and a factor of 2 for the two dots.  The
combination $C^8_2$ is thus the number of combinations to construct a two-electron wave function out of 8 possible states, and represents the
dimensionality of the representation of the permutation group on the combined two-dot two-spin two-valley SU(2)$\times$SU(2)$\times$SU(2)
Hilbert space.

The singlet and triplet states in a Si DQD can be divided into three branches: states with the same valley composition, comprising wave
functions of the form ($++$) and ($--$), and states with mixed valley composition, comprising wave functions of the form ($+-$). There are thus
in total \textit{six} uncoupled branches, which can be labeled $(++)\tilde{S}$, $(+-)\tilde{S}$, $(--)\tilde{S}$ and $(++)\tilde{T}$,
$(+-)\tilde{T}$, $(--)\tilde{T}$ (see Table \ref{tab:states}.) The spatial parts of the $\pm\pm$ singlet and triplet states are
\begin{equation}
\arraycolsep 0.3 ex
\begin{array}{rl}
\displaystyle \tilde{S}^{LR}_{\pm\pm} = & \displaystyle \frac{1}{\sqrt{2}}\, \big( \tilde{L}_\pm^{(1)} \tilde{R}_\pm^{(2)} + \tilde{L}_\pm^{(2)}
\tilde{R}_\pm^{(1)} \big) \\ [1ex] \displaystyle \tilde{S}^{DD}_{\pm\pm} = & \displaystyle \tilde{D}_\pm^{(1)} \tilde{D}_\pm^{(2)} \\ [1ex]
\displaystyle \tilde{T}^{LR}_{\pm\pm} = & \displaystyle \frac{1}{\sqrt{2}}\, \big( \tilde{L}_\pm^{(1)} \tilde{R}_\pm^{(2)} - \tilde{L}_\pm^{(2)}
\tilde{R}_\pm^{(1)} \big).
\end{array}
\end{equation}
For the $+-$ states, the spatial parts of the singlets are
\begin{equation}
\arraycolsep 0.3 ex
\begin{array}{rl}
\displaystyle \tilde{S}^{LR}_{+-} = & \displaystyle \frac{1}{\sqrt{2}}\, \big( \tilde{L}_+^{(1)} \tilde{R}_-^{(2)} + \tilde{L}_+^{(2)}
\tilde{R}_-^{(1)} \big)
\\ [1ex]
\displaystyle \tilde{S}^{LR}_{-+} = & \displaystyle \frac{1}{\sqrt{2}}\, \big( \tilde{L}_-^{(1)} \tilde{R}_+^{(2)} + \tilde{L}_-^{(2)}
\tilde{R}_+^{(1)} \big)
\\ [1ex]
\displaystyle \tilde{S}^{DD}_{+-} = & \displaystyle \frac{1}{\sqrt{2}}\, \big( \tilde{D}_+^{(1)} \tilde{D}_-^{(2)} + \tilde{D}_+^{(2)}
\tilde{D}_-^{(1)} \big),
\end{array}
\end{equation}
while the spatial parts of the triplet states are
\begin{equation}
\arraycolsep 0.3 ex
\begin{array}{rl}
\displaystyle \tilde{T}^{LR}_{+-} = & \displaystyle \frac{1}{\sqrt{2}}\, \big( \tilde{L}_+^{(1)} \tilde{R}_-^{(2)} - \tilde{L}_+^{(2)}
\tilde{R}_-^{(1)} \big)
\\ [1ex]
\displaystyle \tilde{T}^{LR}_{-+} = & \displaystyle \frac{1}{\sqrt{2}}\, \big( \tilde{L}_-^{(1)} \tilde{R}_+^{(2)} - \tilde{L}_-^{(2)}
\tilde{R}_+^{(1)} \big)
\\ [1ex]
\displaystyle \tilde{T}^{DD}_{+-} = & \displaystyle \frac{1}{\sqrt{2}}\, \big( \tilde{D}_+^{(1)} \tilde{D}_-^{(2)} - \tilde{D}_+^{(2)}
\tilde{D}_-^{(1)} \big).
\end{array}
\end{equation}
Because we are using the Wannier states $\tilde{D}_\pm$, there are no overlaps such as $\dbkt{\tilde{L}\pm}{\tilde{R}_\pm}$.  The spatially
symmetric functions $\tilde{S}$ split into three single-valley states from the lower-energy $--$ branch, three single-valley states from the
higher-energy $++$ branch, and four states from the $+-$ branch. The spatially antisymmetric functions $\tilde{T}$ split into one single-valley
state from the lower-energy $--$ branch, another single-valley state from the higher-energy $++$ branch and four states from the $+-$ branch.
\begin{table}[tbp]
  \caption{\label{tab:states} Number of singlet and triplet states in different branches of the spectrum in LDV and ST schemes.}
  $\arraycolsep 1em
   \begin{array}{c@{\hspace{2em}}ccccc} \hline\hline
   & ++  & +-  & -- & {\rm Total} \\ \hline
{\rm LDV \,\, singlet \,\, states} & 3 & 4 & 3 & 10 \\
{\rm LDV \,\, triplet \,\, states} & 3 & 12 & 3 & 18 \\ \hline
{\rm ST \,\, singlet \,\, states} & 2 & 3 & 2 & 7 \\
{\rm ST \,\, triplet \,\, states} & 3 & 9 & 3 & 15 \\ \hline
  \end{array}$
\end{table}
As we discussed before, the smooth quantum dot confinement potential and the electron-electron Coulomb interaction do not introduce further
transitions between valleys (the matrix elements are mostly exponentially suppressed, and the only Coulomb matrix element that is not
exponentially suppressed is nonetheless very small, as we show in the Appendix).  Thus the $--$, $++$ and $+-$ branches do not mix, allowing us
to analyze them separately.

The two-electron Hamiltonians in the $++/--$ branches, in the bases $\{\tilde{S}^{LR}_{\pm\pm}, \tilde{S}^{RR}_{\pm\pm},
\tilde{S}^{LL}_{\pm\pm}, \tilde{T}^{LR}_{\pm\pm} \}$, have the form
\begin{equation}\label{HM}
\begin{array}{rl}
\displaystyle H_{\pm\pm} = & \displaystyle  2\tilde{\varepsilon}_0 \pm 2 |\tilde{\Delta}| + \begin{pmatrix}
\tilde{k} + \tilde{j} & \tilde{t} \sqrt{2} & \tilde{t} \sqrt{2} & 0 \cr
\tilde{t} \sqrt{2} & \tilde{u} & \tilde{j} & 0 \cr
\tilde{t} \sqrt{2} & \tilde{j} & \tilde{u} & 0 \cr
0 & 0 & 0 & \tilde{k} - \tilde{j} \cr
\end{pmatrix}.
\end{array}
\end{equation}
All terms are given in the Appendix. In the $+-$ branch,
\begin{equation}
H^{S, T}_{+-} =  2\tilde{\varepsilon}_0 + \begin{pmatrix}
\tilde{k} & \pm\tilde{j} & \tilde{t} & \tilde{t} \cr
\pm\tilde{j} & \tilde{k} & \pm\tilde{t} & \pm\tilde{t} \cr
\tilde{t} & \pm\tilde{t} & \tilde{u} & \pm\tilde{j} \cr
\tilde{t} & \pm\tilde{t} & \pm\tilde{j} & \tilde{u} \cr
\end{pmatrix},
\end{equation}
where $+$ applies to $H^S_{+-}$ and $-$ to $H^T_{+-}$, and $H^S_{+-}$ is in the basis $\{\tilde{S}^{LR}_{+-}, \tilde{S}^{LR}_{-+},
\tilde{S}^{RR}_{+-}, \tilde{S}^{LL}_{+-} \}$ while for $H^T_{+-}$ replace $\tilde{S} \rightarrow \tilde{T}$. Since the overlap between states
from different valleys is negligible, matrix elements of the form $\tbkt{\tilde{S}_{+ -}}{H_0}{\tilde{S}_{+-}}$ and $\tbkt{\tilde{T}_{+
-}}{H_0}{\tilde{T}_{+-}}$ are equal up to a sign (we will discuss below the case when this condition is not satisfied.) The eigenvalues of
$H^S_{+-}$ and $H^T_{+-}$ are exactly the same. In other words, there is no exchange splitting between the singlet and triplet states if the two
electrons are in different valley eigenstates.

\subsection{Scalability}

The structure of the two-electron Hamiltonian has important consequences for the operation of spin qubits in silicon.  The $++$ and $--$
branches have the same structure as the single-valley Hamiltonian except they are shifted up and down by $2|\tilde{\Delta}|$ respectively. These
branches can be regarded as replicas of the usual single-valley Hund-Mulliken Hamiltonian.\cite{Slater_Molecules, Burkard_PRB99} The singlet and
triplet states have the same form as in the single-valley case.  The principal difference from the one-valley situation emerges when we consider
the $+-$ branches, where we have four sets of degenerate singlet and triplet states.  This degeneracy is the direct manifestation of the fact
that exchange integrals for electrons from different valleys vanish (either exponentially small or very small due to Bloch function symmetry,
see Appendix).

The essence of exchange gates in the Loss-DiVincenzo architecture is the control of the phase difference (introduced by a finite exchange
splitting) between singlet and triplet states.  In the $--$ or $++$ branches the two electrons are in the same valley eigenstate, so that we
exactly recover the single-valley physics of GaAs, and exchange gates can be implemented.  However, in the $+-$ and $-+$ branches singlet and
triplet states are degenerate, rendering exchange gates impossible.  Now assume that the two electrons were initialized into two initially
independent quantum dots with an arbitrary combination of valley eigenstates.  When the inter-dot barrier is lowered, if $|\tilde{\Delta}|$ is
sufficiently small, the two-electron states would generally populate all four branches ($++$, $+-$, $-+$, and $--$).  Since exchange gates
cannot be implemented in the $+-$ and $-+$ branches, they cannot be implemented for these two electrons in general.  The only way to perform
exchange gates is to have the two-electron states populating only the $++$ and $--$ branches.  With the electrons in a reservoir in a
structure-less Fermi distribution, the only single-electron state that can be loaded exclusively is the $-$ valley eigenstate (if
$2|\tilde{\Delta}| \gg k_B T$), so that when inter-dot tunnelling is turned on, the two electrons are in the $--$ branch.  This is the only
viable option where exchange gates can still be performed. The exchange splitting between the lowest-lying singlet and triplet states in the
$--$ branch is given by the Hund-Mulliken expression,\cite{Burkard_PRB99, Qiuzi_PRB10} which can be obtained by straightforward diagonalization
of Eq.\ (\ref{HM}),
\begin{equation}
\tilde{J} = - 2 \tilde{j} - \frac{(\tilde{u} - \tilde{k})}{2} + \frac{1}{2}\, \sqrt{(\tilde{u} - \tilde{k})^2 + 16 \tilde{t}^2}.
\end{equation}
In the $++$ and $- -$ branches there are four energy levels: a singlet, a triplet which is higher by $\tilde{J}$, and two singlets which are
much higher in energy due to the on-site Coulomb interaction. In the $+-$ branch one has the same four levels (bar the valley splitting), except
that for each energy there is a singlet and a triplet. In this branch exchange is zero, and the next highest state is separated by $\tilde{J}$.
Therefore exchange in the three branches is either $\tilde{J}$ or zero. This fact becomes important when $|\tilde{\Delta}| \le k_BT$. Firstly,
for $\tilde{\Delta} = 0$, the lowest state is a (sixfold) degenerate singlet/triplet. For $\tilde{\Delta}$ nonzero but $|\tilde{\Delta}| \le
k_BT$, one may in general initialize a superposition of states and exchange is not well defined. Consider two qubits $(\alpha_L \tilde{L}_+ +
\beta_L \tilde{L}_-)$ and $(\alpha_R \tilde{R}_+ + \beta_R \tilde{R}_-)$. If $\alpha_L = \beta_R = 1$ the states are $\tilde{L}_+$ and
$\tilde{R}_-$, and the exchange coupling between them is zero. For arbitrary $\alpha_D$, $\beta_D$ exchange depends on the specific values of
$\alpha_D$ and $\beta_D$ and cannot be controlled, precluding QC operations.

Our analysis shows that in the LDV architecture for multivalley systems single-qubit operations are straightforward, while exchange-based
two-qubit operations are not feasible unless $|\tilde{\Delta}| \gg k_BT$. The optimal strategy is to strive to obtain a large valley splitting
and work towards replicating the single-valley situation.

\section{Singlet-triplet qubits}
\label{sec:ST}

We have studied the implications of multiple valleys on the implementation of singlet-triplet qubits in Ref.\ \onlinecite{Culcer_PRB09}.  In
this Section, aside from reviewing initialization (Sec. \ref{sec:init}), we provide the important technical details and explore new features
such as the evolution of the spectrum as a function of magnetic field, intravalley and intervalley effects due to the hyperfine interaction,
the role of these terms in the dynamics of singlet-triplet qubits, the difference between applied and hyperfine magnetic fields, spin blockade
and two-qubit operations.

We use the notation ($n$,$m$) to indicate the occupancy of the left and right dots respectively. The single-particle energies on the left and
right dots are different, $\tilde{\varepsilon}_R = \tbkt{\tilde{R}_\xi}{(T+V_Q)}{\tilde{R}_\xi}$ and $\tilde{\varepsilon}_L =
\tbkt{\tilde{L}_\xi}{(T+V_Q)}{\tilde{L}_\xi}$. The dimensionless detuning is defined as $\delta = (\tilde{\varepsilon}_L -
\tilde{\varepsilon}_R)/(2d\tilde{\varepsilon}_0)$, and the dimensionless critical detuning $\delta_c = (\tilde{u} -
\tilde{k})/(2d\tilde{\varepsilon}_0)$.

\begin{figure}[tbp]
\begin{tabular}{lr}
\includegraphics[width=0.9\columnwidth]{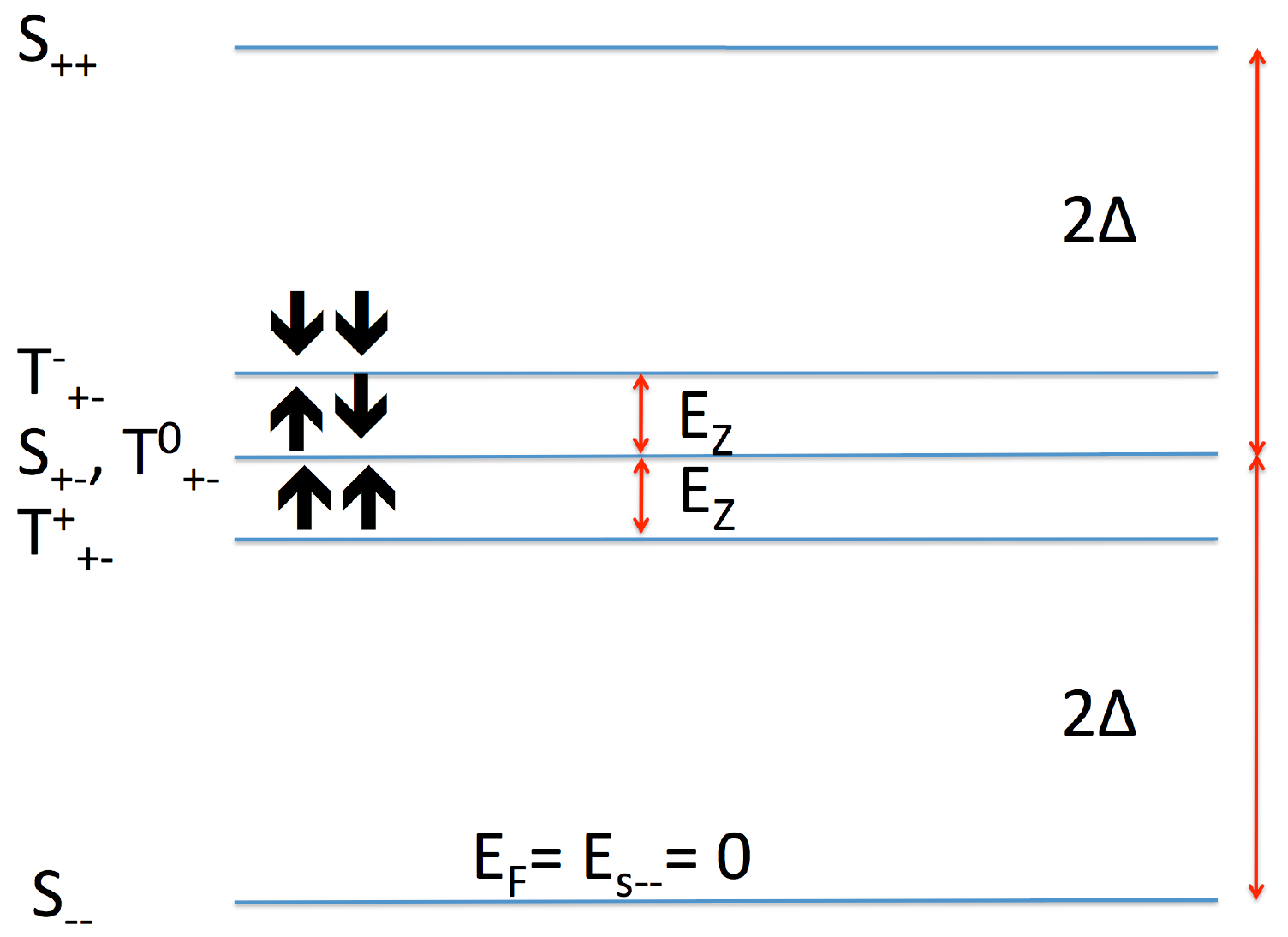}
\end{tabular}
\caption{Two-electron energy levels on a single dot at finite $\Delta$ and magnetic field such that $2|\Delta| > E_Z$ as introduced in
Ref.~\onlinecite{Culcer_PRB09}. In this case the lowest energy state is the singlet $S_{--}$, followed by the triplet $T^{\uparrow
\uparrow}_{+-}$, the degenerate singlet $S_{+-}$/triplet $T^{\uparrow \downarrow}_{+-}$ and triplet $T^{\downarrow \downarrow}_{+-}$, and
finally the singlet $S_{++}$. Electron spin orientations in $T$ are indicated by arrows.} \label{SingleDotLevels}
\end{figure}

\subsection{Initialization}
\label{sec:init}

Our study of the singlet-triplet qubit follows the successful experimental implementation in GaAs DQDs, which has 4 basic ingredients:
initialization in a single dot in the (0,2) regime, single-qubit manipulation in the (1,1) regime, two-qubit manipulation via electrostatic
interaction, and spin measurement via spin blockade. \cite{Petta_Science05, Taylor_PRB07}

First a (0,2) state is initialized ($E > 0$.) Using the notation of Sec.\ \ref{sec:Loss_PRA98}, the four lowest-energy two-particle spatial wave
functions are,
\begin{equation}\label{SingleDot}
\arraycolsep 0.3 ex
\begin{array}{rl}
\displaystyle \tilde{S}^{RR}_{\pm \pm} = & \displaystyle \tilde{R}_\pm^{(1)} \tilde{R}_\pm^{(2)}
\\ [1ex]
\displaystyle \tilde{S}^{RR}_{+-} = & \displaystyle (1/\sqrt{2}) \big(\tilde{R}_+^{(1)} \tilde{R}_-^{(2)}  + \tilde{R}_+^{(2)}
\tilde{R}_-^{(1)}\big)
\\ [1ex]
\displaystyle \tilde{T}^{RR}_{+-} = & \displaystyle(1/\sqrt{2}) \big(\tilde{R}_+^{(1)} \tilde{R}_-^{(2)} - \tilde{R}_+^{(2)}
\tilde{R}_-^{(1)}\big),
\end{array}
\end{equation}
In the basis $\{ \tilde{S}^{RR}_{- -}, \tilde{S}^{RR}_{+ -}, \tilde{T}^{RR}_{+ -} , \tilde{S}^{RR}_{++} \}$ the Hamiltonian is
\begin{equation}\label{1dot}
H_{1dot} = 2 \tilde{\varepsilon}_0 + \tilde{u} + \begin{pmatrix}
-2|\tilde{\Delta}| & 0 & 0 & 0 \cr
0 & 0 & 0 & 0 \cr
0 & 0 & 0 & 0 \cr
0 & 0& 0 & 2|\tilde{\Delta}|
\end{pmatrix}.
\end{equation}
For $\tilde{\Delta} = 0$ and no external magnetic field ${\bm B}$, all six lowest-energy levels are degenerate and it is not possible to load
any particular two-electron state. An external magnetic field splits the triplet energy levels by the Zeeman energy $E_Z$
(Fig.~\ref{SingleDotLevels}.) Initialization of a (0,2) state requires an outside reservoir with a Fermi energy $\varepsilon_F$ (thermally
broadened by $\approx k_BT$) tuned to be on resonance with $\tilde{S}^{RR}_{--}$. If $|\tilde{\Delta}| \gg k_BT$ (at T=100mK we require
$\tilde{\Delta} \approx$ 0.1meV) then $\tilde{S}^{RR}_{- -}$ can be loaded exclusively.

The qubit of Refs.~\onlinecite{Petta_Science05, Hanson_RMP07} is in a single-valley system, so $\tilde{R}_\pm \rightarrow \tilde{R}$.  Only a
singlet can be made out of the lowest lying wave function. The triplet involves higher orbitals and is separated by $\varepsilon_0 \gg k_BT$,
minus a small exchange term. In GaAs this energy scale is $\approx$ 1 meV, which is much greater than the experimental thermal energy of 0.01
meV (100mK). In Si the energy scale is set by $\tilde{\Delta}$. If $\tilde{\Delta}$ is sufficiently large the physics is that of a single-valley
problem.

In the evaluation of Eq.\ (\ref{1dot}) we have also encountered a \textit{valley-exchange} Coulomb integral
\begin{equation}
j_v = \int d^3 r_1 \int d^3 r_2\, \tilde{D}_z^{*(1)} \tilde{D}_{\bar{z}}^{*(2)} V_{ee} \, \tilde{D}_{\bar{z}}^{(1)} \tilde{D}_z^{(2)}.
\end{equation}
This integral is not suppressed by an exponentially small prefactor, yet it is $\ll 1\mu$eV and we do not take it into account.  Its evaluation
is discussed in Appendix \ref{sec:valleyx}.

\subsection{One-qubit manipulation}
\label{sec:onequbit}

Single-qubit operations on the singlet-triplet qubit are performed in the (1,1) regime.\cite{Petta_Science05}  The two-particle states are the
same as those enumerated in the case of single-spin qubits, except here we do not include the high-energy (2,0) states of the form
$\tilde{L}\tilde{L}$ (see Table\ \ref{tab:states}.) In the bases $\{\tilde{S}^{LR}_{\pm\pm}, \tilde{S}^{RR}_{\pm\pm}, \tilde{T}^{LR}_{\pm\pm}
\}$ the matrix elements of the Hamiltonian in the $++/--$ branches are given by
\begin{equation}
H_{\pm\pm} =  \tilde{\varepsilon}_L + \tilde{\varepsilon}_R + \tilde{k} \pm 2|\tilde{\Delta}| + \begin{pmatrix}
\tilde{j} & \tilde{t}\sqrt{2} & 0 \cr
\tilde{t}\sqrt{2} & - 2d\varepsilon_0 (\delta - \delta_c) & 0 \cr
0 & 0 & - \tilde{j} \cr
\end{pmatrix}.
\end{equation}
The matrix elements in the $+-$ branch, in the bases $\{\tilde{S}^{LR}_{+-}, \tilde{S}^{LR}_{-+}, \tilde{S}^{RR}_{+-} \}$ and
$\{\tilde{T}^{LR}_{+-}, \tilde{T}^{LR}_{-+}, \tilde{T}^{RR}_{+-} \}$, are
\begin{equation}
H^{S,T}_{+-} =  \tilde{\varepsilon}_L + \tilde{\varepsilon}_R + \tilde{k} + \begin{pmatrix}
0 & \pm \tilde{j} & \tilde{t} \cr
\pm \tilde{j} & 0 & \pm \tilde{t} \cr
\tilde{t} & \pm \tilde{t} & - 2d\varepsilon_0 (\delta - \delta_c) \cr
\end{pmatrix},
\label{eq:HS+-}
\end{equation}
where the $\pm$ signs apply to the singlet/triplet Hamiltonians respectively.

The $+-$ singlet and triplet branches always yield the same energies. Moreover, for $\tilde{\Delta}=0$ the lowest energy state is sixfold
degenerate: the lowest energy $++$, $+-$ and $--$ singlets, plus the lowest energy $+-$ triplet. When the two electrons are in the \textit{same}
valley, the singlet and triplet energies differ by $\tilde{J}$.

For clarity we focus on a concrete example, taking a Si DQD with a=8.2nm, b=3nm, d=2.45 and $|\tilde{\Delta}|$=0.1meV. The energy levels of the
system with these parameters are plotted in Fig.~\ref{LargeDelta_All} (analogous to Ref.\ \onlinecite{Culcer_PRB09}) as a function of the
dimensionless detuning $\delta$. At low detuning there are four (0,2) high-energy levels, indicated by the two solid lines (representing
singlets of the form $\tilde{R}_+\tilde{R}_+$, $\tilde{R}_-\tilde{R}_-$) and one dashed line (representing two degenerate singlet and triplet of
the form $\tilde{R}_+\tilde{R}_-$). The separation of these levels is 2$|\tilde{\Delta}|$.  The lower-energy (1,1) levels are a degenerate
singlet/triplet of the form $\tilde{L}_+\tilde{R}_+$ (top solid line), a degenerate singlet/triplet of the form $\tilde{L}_-\tilde{R}_-$ (bottom
solid line), and two degenerate valley mixing singlets and triplets of the form $\tilde{L}_+ \tilde{R}_-$ and $\tilde{L}_-\tilde{R}_+$. For the
parameters considered in this example $\tilde{t} \approx$ 0.02meV,  giving a splitting at the avoided crossing of $\approx$ 0.06meV.

\subsubsection{Double dot spectra in different parameter regimes}
\label{sec:spectra}

\begin{figure}[t]
\includegraphics[width=\columnwidth]{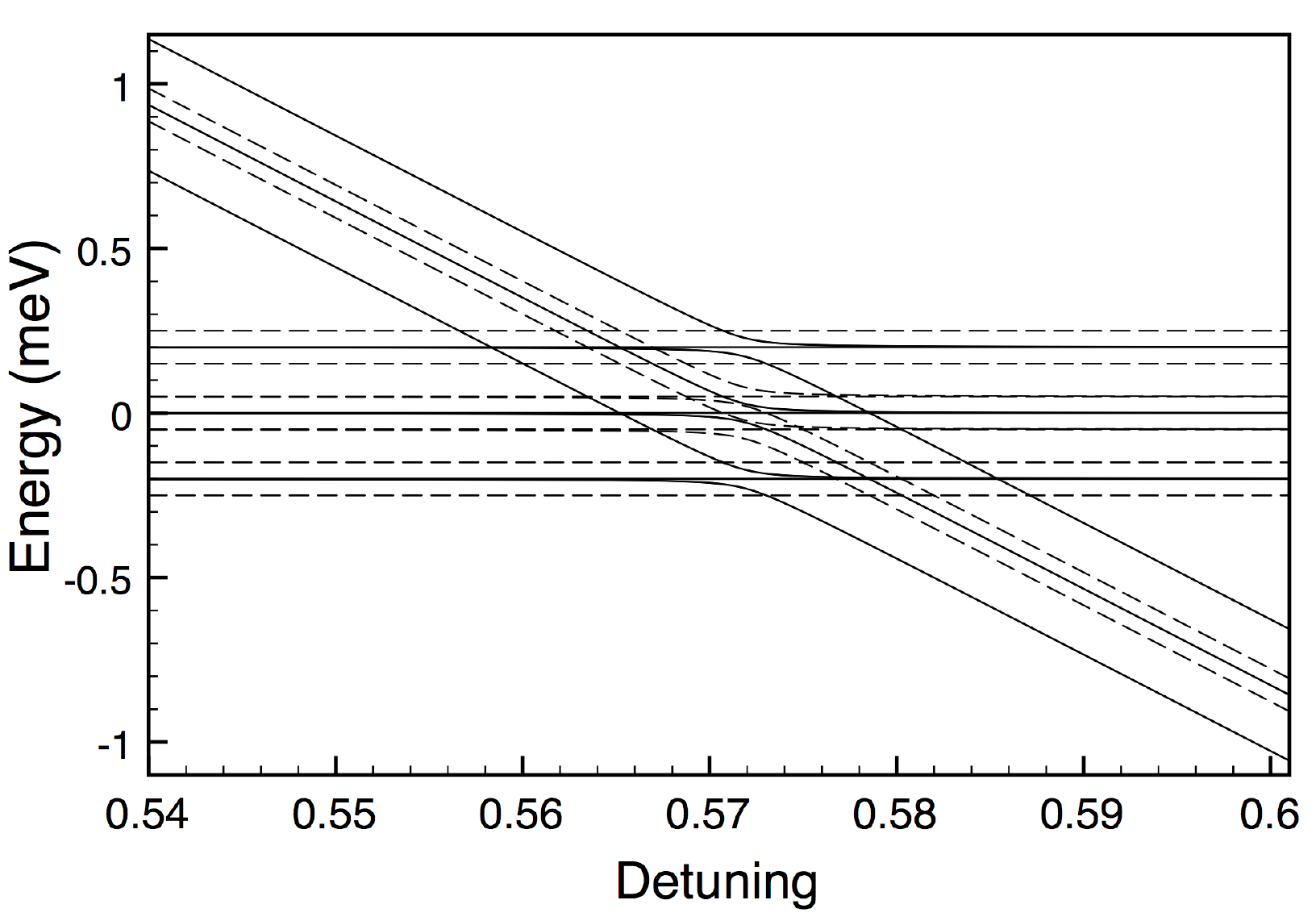}
\caption{Energy level spectrum for a Si/SiO$_2$ DQD with a=8.2nm, b=3nm, d=2.45 and $|\tilde{\Delta}|$=0.1meV, $E_Z$ = 0.05 meV,
$\tilde{t}$=0.02meV. The solid lines represent singlet and unpolarized triplet levels $\tilde{S}$ and $\tilde{T}^0$, while the dashed lines
represent polarized triplet levels $\tilde{T}^+$ and $\tilde{T}^-$.  The top and bottom anticrossings each consist of two singlets and one
triplet In the middle anticrossing each of the three dashed lines represents a degenerate singlet/triplet level.} \label{LargeDelta_All}
\end{figure}

\begin{figure}[t]
\includegraphics[width=\columnwidth]{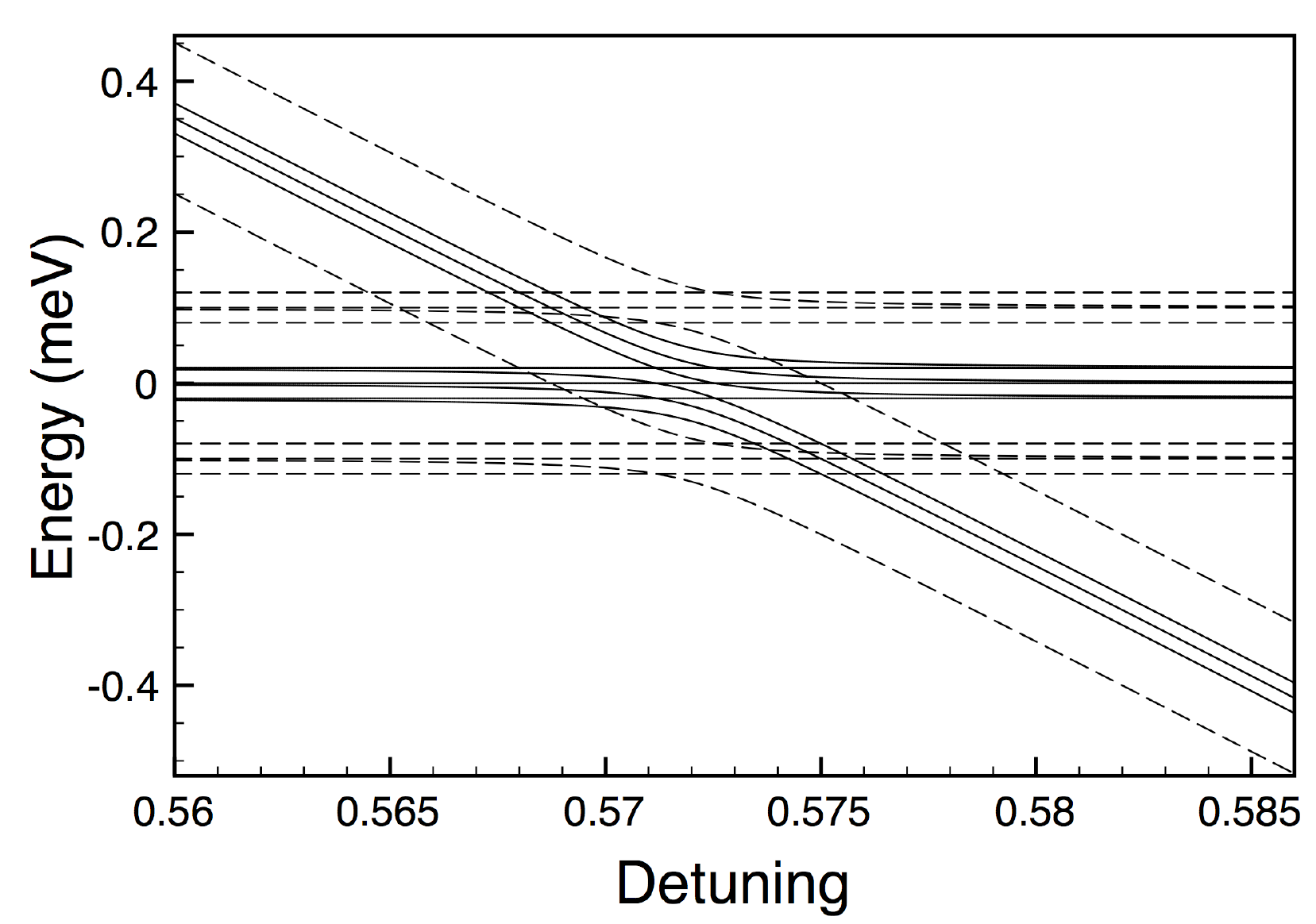}
\caption{Energy level spectrum for a Si/SiO$_2$ DQD with a=8.2nm, b=3nm, d=2.45 and $|\tilde{\Delta}|$=0.01meV, $E_Z$ = 0.1 meV,
$\tilde{t}$=0.02meV.  Here the valley splitting $|\tilde{\Delta}|$ has been set as the smallest energy scale, the Zeeman energy as the largest
scale, and the tunnel coupling in between. This figure illustrates the opposite scenario to Fig. \ \ref{LargeDelta_All}, showing the
qualitatively different structure of the energy spectrum when $|\tilde{\Delta}| < \tilde{t}$.} \label{LargeZeeman_All}
\end{figure}

\begin{figure}[t]
\includegraphics[width=\columnwidth]{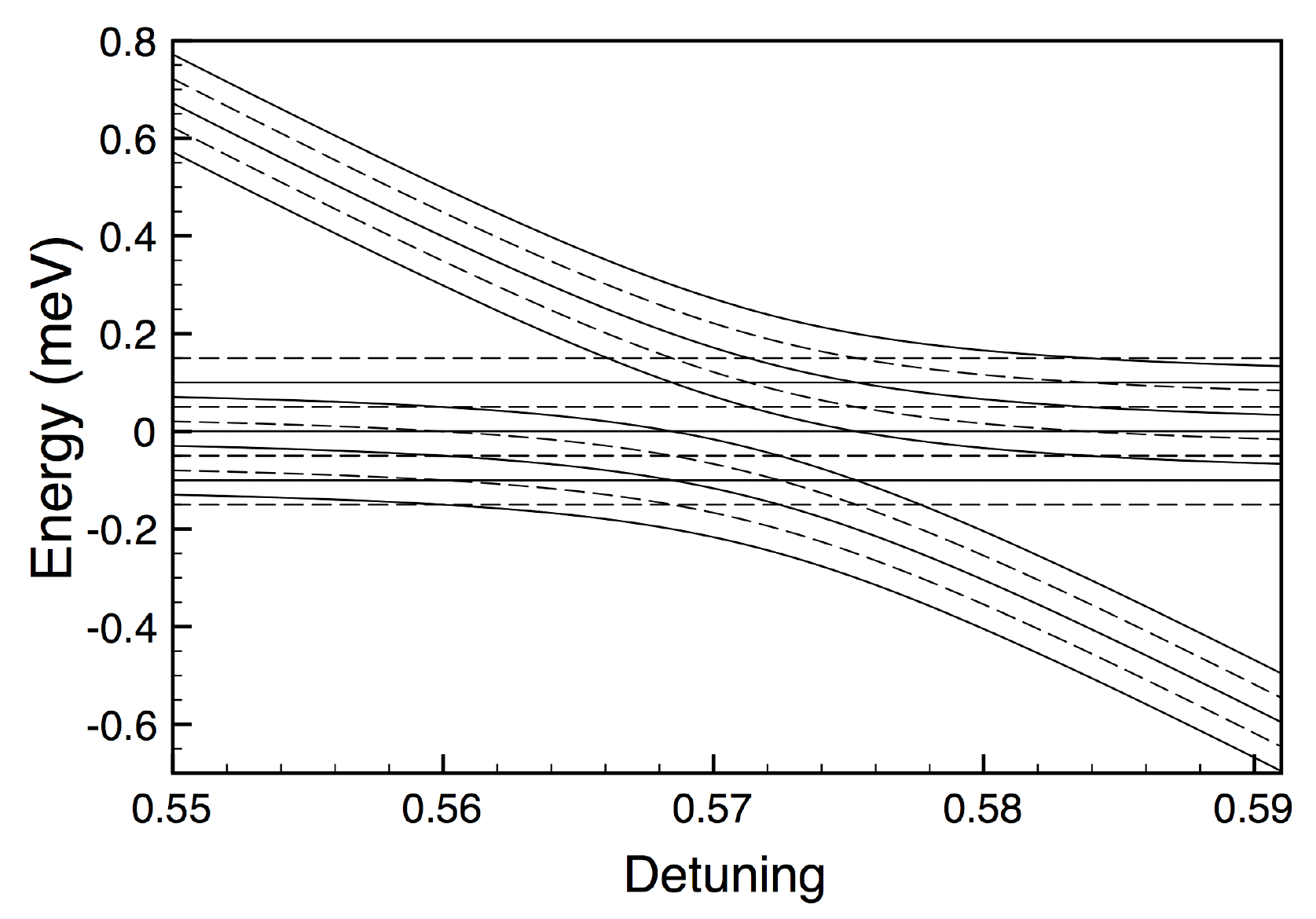}
\caption{Energy level spectrum for a Si/SiO$_2$ DQD with a=8.2nm, b=3nm, d=2.45 and $|\tilde{\Delta}|$=0.05meV, $E_Z$ = 0.05 meV,
$\tilde{t}$=0.1meV.  In this figure the tunnel coupling has been set as the largest energy scale. Although the magnitude of $\tilde{t}$ in this
graph has been exaggerated for clarity and exceeds what one expects to measure experimentally, this figure illustrates the complications
inherent in experiments seeking to distinguish parameters of comparable magnitude.} \label{Intermed_All}
\end{figure}

\begin{figure}[t]
\includegraphics[width=\columnwidth]{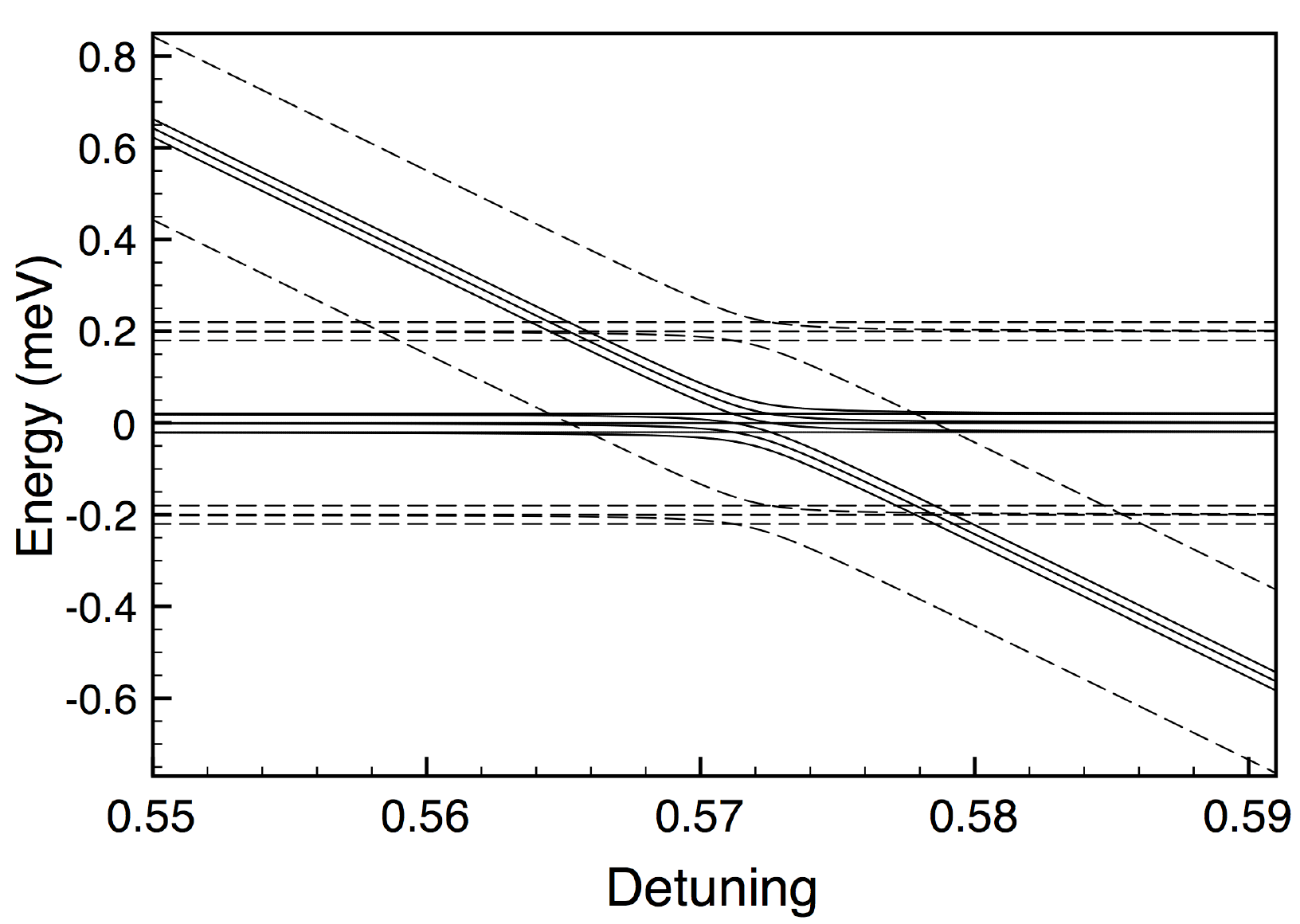}
\caption{Energy level spectrum for a Si/SiO$_2$ DQD with a=8.2nm, b=3nm, d=2.45 and $|\tilde{\Delta}|$=0.01meV, $E_Z$ = 0.2 meV,
$\tilde{t}$=0.02meV.  In this figure the Zeeman energy has been set as the largest energy scale and the valley splitting as the lowest. Notice
the qualitative similarity of this figure to Fig.\ \ref{LargeDelta_All}. As the magnetic field increases, in effect the Zeeman field and the
valley-orbit coupling trade places.} \label{HugeZeeman_All}
\end{figure}

\begin{figure}[t]
\includegraphics[width=\columnwidth]{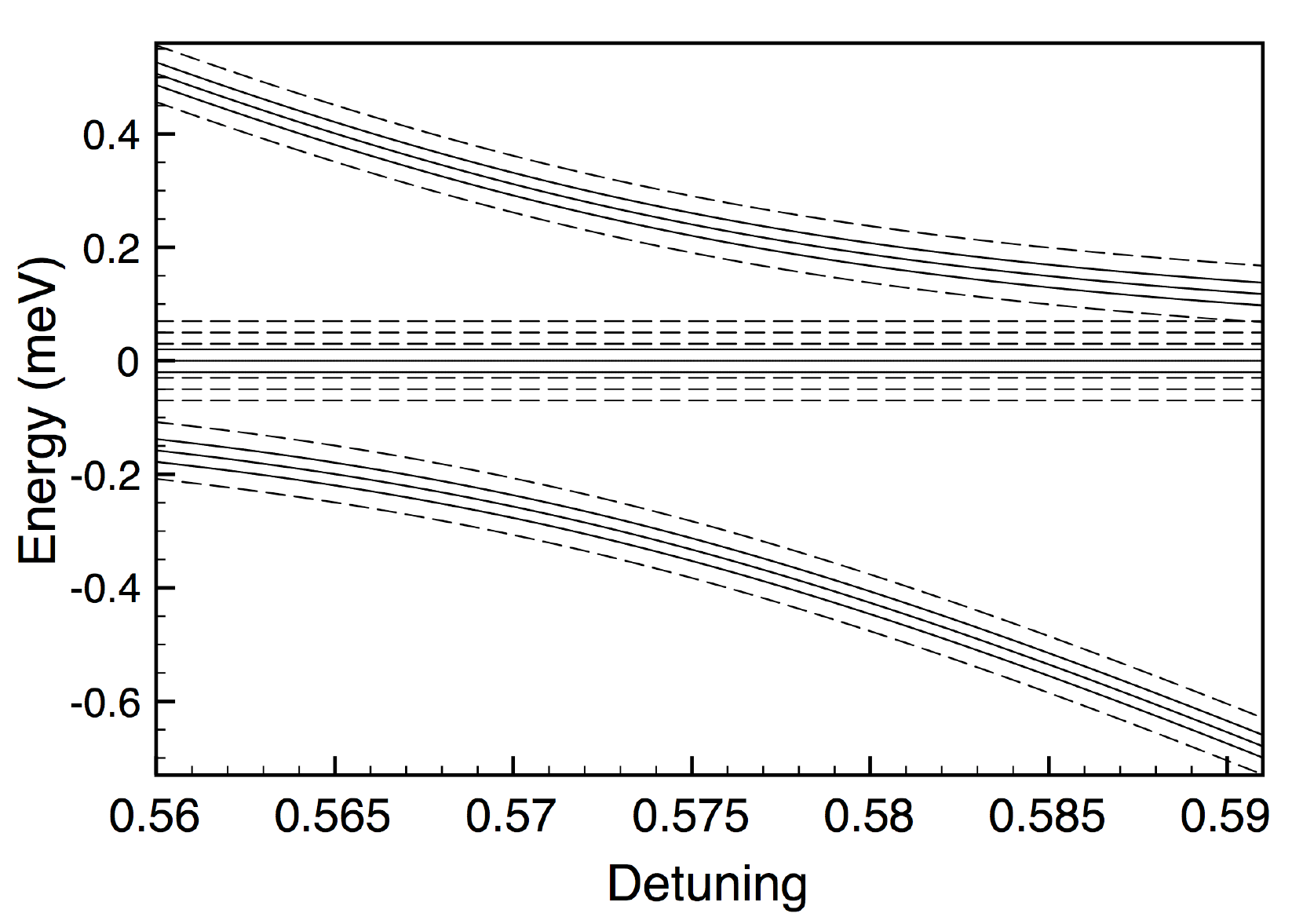}
\caption{Energy level spectrum for a Si/SiO$_2$ DQD with a=8.2nm, b=3nm, d=2.45 and $|\tilde{\Delta}|$=0.01meV, $E_Z$ = 0.05 meV,
$\tilde{t}$=0.2meV.  This figure illustrates the case when the tunnel coupling is the largest energy scale in the problem.}
\label{SmallDelta_All}
\end{figure}

Up to now we have considered the simplest form of the two-dot spectrum, which occurs in the case when $|\tilde{\Delta}|$ is the largest energy
scale,  exceeding the tunnel coupling and the Zeeman splitting due to the uniform magnetic field by a noticeable amount. Yet it is evident that
the qualitative features of the spectrum are sensitively dependent on the relative size of $\tilde{t}$, $|\tilde{\Delta}|$, and $E_Z$. As these
parameters vary with respect to one another, the relative position of most energy levels can differ greatly. As a result, there can be
substantial variation in the loading and mixing dynamics of the two-electron states (Figs.\ \ref{LargeDelta_All}--\ref{SmallDelta_All}.)

In Fig. \ref{LargeDelta_All} we plot the two-electron spectrum of a Si DQD when $|\tilde{\Delta}| \gg \tilde{t}$ including \textit{all} the
Zeeman-split levels. It is interesting to follow the evolution of the two-dot two-electron spectrum as the ratio of the magnetic field to the
valley splitting goes from small to large. This can be done by observing the way the spectrum changes from Figs. \ref{LargeDelta_All} through
\ref{HugeZeeman_All}. These figures taken together illustrate the fact that, as the magnetic field increases with respect to the valley coupling
there is a transition between the valley physics (which is in effect a pseudospin) and the Zeeman physics (which is due to the real spin). For
large valley/Zeeman splitting the spectrum looks effectively the same, except in one extreme case the splitting between the three sets of levels
is determined by $|\tilde{\Delta}|$ whereas in the other extreme case it is determined by $E_Z$.

On the other hand Fig. \ref{SmallDelta_All} contains the two-dot spectrum for the opposite case when $|\tilde{\Delta}| \ll \tilde{t}$. In
Figs.~\ref{LargeDelta_All} and \ref{SmallDelta_All}  we have assumed the same value of the Zeeman splitting. Finally, in general case one should
be prepared for an intermediate scenario as in Fig.~\ref{Intermed_All} in which the sets of levels are not clearly separated and may cross. In
such a situation distinguishing the energy levels experimentally may prove challenging.

\subsubsection{Inhomogeneous magnetic fields in Si}

Quantum coherent single-qubit experiments on singlet-triplet qubits make use of an inhomogeneous magnetic field. As shown in
Ref.~\onlinecite{Petta_Science05},  an experiment on the singlet-triplet qubit often requires two magnetic fields: a uniform one to separate out
the polarized triplet states, and an inhomogeneous one to mix the singlet and the unpolarized triplet states. In the existing experiments on
GaAs\cite{Petta_Science05} the latter is provided by the lattice nuclear spins.\cite{Foletti_NP09} In Si, hyperfine interaction is much weaker,
and number of lattice nuclear spins much smaller.\cite{Assali_preprint10} Natural Si has only $4.7\%$ of the $^{29}$Si isotope having nonzero
nuclear spin ($^{28}$Si has no nuclear spin) and it can be further isotopically purified to nearly completely eliminate the nuclear spins. For
the case when the inhomogeneous effective field arises from hyperfine interaction with the nuclei in the dots, the magnetic fields on the left
and right dots ($\bm B_L$ and $\bm B_R$ below) differ by a small amount of the order of $\mathcal{A}/\sqrt{N}$ where $\mathcal{A}$ is the
hyperfine constant of the material and $N$ is the number of nuclei interacting appreciably with the electrons. In gated GaAs QDs this amount is
on the order of mT. In a natural Si QD it has been estimated to be in the range 0.06-6 neV, corresponding to a time scale of 0.1-10$\mu$s and
magnetic fields of 0.5-50 $\mu$T, depending on the percentage of $^{29}$Si in the material, from 0.01$\%$ to 100$\%$.\cite{Assali_preprint10}
In systems based on purified Si, an external inhomogeneous magnetic field must be generated, such as by a nanomagnet,\cite{Pioro_NatPhys08}
allowing better control.  For example, one can tailor the inhomogeneous field to have only a $\hat{\bm z}$-component with a gradient along the
$\hat{\bm x}$-direction.

Each initialized (0,2) state has a probability of return to the original state after staying for a certain time in the (1,1) regime.  In a
random hyperfine field this probability of return is determined by the nuclear spin polarization difference between the two dots.  Since the
experiment is repeated many times in the time domain, the observed results involve averages over the inhomogeneous nuclear field.  On the other
hand, in an applied inhomogeneous magnetic field, the field is the same during each experimental run. Therefore, by controlling the magnitude
and orientation of the applied inhomogeneous field one can control the composition of the (1,1) state after mixing, and therefore the return
probability. In this case, if the initial state is known, we have a controlled return probability. The key fact is that, in the far-detuned
(1,1) regime where $\delta \ll \delta_c$, each unpolarized state (singlet or triplet) mixes with one other unpolarized state, while polarized
triplets do not mix with any other state (see Sec.~\ref{sec:mixing}). Knowledge of the controlled return probability requires exact knowledge
and control of the applied inhomogeneous magnetic field inside each dot.

\subsubsection{Singlet-triplet mixing in an applied inhomogeneous magnetic field}
\label{sec:mixing}

The inhomogeneous magnetic field is the sum of nuclear and applied fields, ${\bm B}({\bm r}) = {\bm B}_{nuc}({\bm r}) + {\bm B}_{appl}({\bm
r})$.  Experimentally one cannot generate an applied field with a profile sufficiently sharp in the $\hat{\bm z}$-direction to mix states from
different valleys, so for ${\bm B}_{appl}({\bm r})$ the Zeeman Hamiltonian does not mix $z$ and $\bar{z}$ states. In the case of the nuclear
field, the contact hyperfine interaction is a sum of $\delta$-functions, and can mix states from different valleys. Here we consider ${\bm
B}({\bm r}) = {\bm B}_{appl}({\bm r})$. Considerations specific to ${\bm B}_{nuc}({\bm r})$ are discussed in the following section.

The Zeeman Hamiltonian in Si, where $g = 2$ approximately, for two electrons in a DQD is
\begin{equation}
H_Z = - \mu_B \, [ {\bm \sigma}_1 \cdot {\bm B}({\bm r}_1) + {\bm \sigma}_2 \cdot {\bm B}({\bm r}_2)]
\end{equation}
where the spin operators ${\bm \sigma}_{1,2}$ act on the spin of particle 1 and 2 respectively.  We study the Zeeman Hamiltonian in the subspace
of two-electron (1,1) states formed out of the wave functions $\{  \tilde{D}_\pm \}$, which is relevant to operations on singlet triplet qubits.
Its expectation values contain expressions of the form (assuming weak tunnel coupling)
\begin{equation}\label{InhomoB}
\arraycolsep 0.3 ex
\begin{array}{rl}
\displaystyle \tbkt{\tilde{D}_\xi}{{\bm B}({\bm r})}{\tilde{D}_\xi} = & \displaystyle \bm B_D \\ [3ex]
\displaystyle \tbkt{\tilde{L}_\xi}{{\bm B}({\bm r})}{\tilde{R}_\xi} \approx & \displaystyle 0.
\end{array}
\end{equation}
We write these (real or effective) magnetic fields in the form $\bm B_{L, R} = (\bm B_{tot} \pm \Delta \bm B)/2$, where the total magnetic field
is $\bm B_{tot} = \bm B_L + \bm B_R$ and the field difference is $\Delta \bm B = \bm B_L - \bm B_R$. Here, since ${\bm B}({\bm r}) = {\bm
B}_{appl}({\bm r})$, the magnetic field is smooth in space and has no matrix elements between states $z$ and $\bar{z}$. In this case, since
$\tbkt{\tilde{D}_{z, \bar{z}}^{(1)}}{{\bm B}({\bm r}_1)}{\tilde{D}_{\bar{z}, z}^{(1)}} \approx 0$, the matrix elements of the Hamiltonian
between the states $\{  \tilde{D}_\pm \}$ are easy to find using Eq.\ (\ref{InhomoB})
\begin{equation}
\arraycolsep 0.3 ex
\begin{array}{rl}
\displaystyle \tbkt{\tilde{D}_\pm}{{\bm B}({\bm r})}{\tilde{D}_\pm} = & \displaystyle \bm B_D \\ [3ex]
\displaystyle \tbkt{\tilde{L}_\pm}{{\bm B}({\bm r})}{\tilde{R}_\pm} \approx & \displaystyle 0.
\end{array}
\end{equation}
Within the mixed-valley branch, neither an applied field nor the nuclear field can mix $+-$ states with $-+$ states in the mean field regime.
We only consider the matrix elements of the Zeeman Hamiltonian within each branch of the two-dot spectrum, i.e. $++$, $--$ and $+-$.

Singlet-triplet mixing occurs in the (1,1) configuration, when $\delta \ll \delta_c$.\cite{Taylor_PRB07, Petta_Science05}  Without putting in
valley eigenstate indices, we are interested in the effect of the Zeeman term in the space of states spanned by a (1,1) singlet $\Psi^{LR}_S$
and a (1,1) triplet $\Psi^{LR}_T$.  An applied uniform magnetic field separates the polarized triplet states energetically, so that the
two-electron system remains in the branch spanned only by $\Psi^{LR}_S$ and $\Psi^{LR}_{T_0}$, where
\begin{equation}\label{Zeeman}
\tbkt{\chi_S}{H_Z}{\chi_S} = \tbkt{\chi_0}{H_Z}{\chi_0} = 0.
\end{equation}
We note that
\begin{equation}
\begin{array}{rl}
\displaystyle \tbkt{\chi_S}{{\bm \sigma}^{(i)}}{\chi_S} = & \displaystyle \tbkt{\chi_{T_0}}{{\bm \sigma}^{(i)}}{\chi_{T_0}} = 0 \\ [3ex]
\displaystyle \tbkt{\chi_S}{{\bm \sigma}^{(1)}}{\chi_{T_0}} = & \displaystyle \hat{\bm z} \\ [3ex]
\displaystyle \tbkt{\chi_S}{{\bm \sigma}^{(2)}}{\chi_{T_0}} = & \displaystyle - \hat{\bm z}.
\end{array}
\end{equation}
and the off-diagonal matrix element is
\begin{equation}
\arraycolsep 0.3 ex
\begin{array}{rl}
\displaystyle \tbkt{\Psi^{LR}_S}{H_Z}{\Psi^{LR}_{T_0}} = - \mu_B  \, \Delta B_z.
\end{array}
\end{equation}
The $x$- and $y$-components of $\Delta{\bm B}$ mix the singlet with the polarized triplets
\begin{equation}
\arraycolsep 0.3 ex
\begin{array}{rl}
\displaystyle \tbkt{\Psi^{LR}_S}{H_Z}{\Psi^{LR}_{T_+}} = \frac{\mu_B}{\sqrt{2}}  \, (\Delta B_x + i \Delta B_y) \\ [3ex]
\displaystyle \tbkt{\Psi^{LR}_S}{H_Z}{\Psi^{LR}_{T_+}} = - \frac{\mu_B}{\sqrt{2}}  \, (\Delta B_x - i \Delta B_y).
\end{array}
\end{equation}
The polarized triplets are split off by the uniform Zeeman field and transitions to them are suppressed.  The triplets are mixed among
themselves by the total magnetic field
\begin{equation}
\arraycolsep 0.3 ex
\begin{array}{rl}
\displaystyle \tbkt{\Psi^{LR}_{T_0}}{H_Z}{\Psi^{LR}_{T_\pm}} = - \frac{\mu_B}{\sqrt{2}}  \, ( B_x \pm i  B_y) \,.
\end{array}
\end{equation}
The uniform Zeeman field also suppresses this mixing.  In an applied uniform magnetic field $\tilde{S}^{LR}_{\pm\pm}$ mix only with
$\tilde{T}^{LR}_{\pm\pm}$, and $\tilde{S}^{LR}_{\pm\mp}$ mix only with $\tilde{T}^{LR}_{\pm\mp}$. Each unpolarized (1,1) state only mixes with
one other unpolarized (1,1) state, which is important in spin blockade.

\subsubsection{The hyperfine interaction in Si}
\label{sec:hfSi}

The hyperfine interaction between electron and nuclear spins in Si is relatively well known.  For donor-bound electrons, the full hyperfine
matrix, including both the contact and anisotropic hyperfine couplings to the donor nuclei and the $^{29}$Si nuclear spins in the environment,
has been carefully measured and calculated.\cite{Feher_PR59, Hale_PR69, Hale_PRB71, Ivey_PRB75} The thorough understanding of that problem
allows quite precise knowledge of the coupled electron-nuclear-spin dynamics in Si:P, as represented by the impressive agreement between theory
and experiment in ESR experiments on Si:P.\cite{Witzel_AHF_PRB07} The hyperfine interaction with conduction electrons in Si is generally known
to be weak relative to confined donor electrons. The earliest experimental measurement of hyperfine coupling strength for conduction electrons
in Si dated back to the 1950s and 60s. \cite{Shulman_PR56,Wilson_PR64,Dyakonov_PRB92}  The contact hyperfine interaction, measured by average
electron probability $\eta = |\psi(0)|^2/\langle \psi^2 \rangle$ at nuclear sites, ranges from 178,\cite{Wilson_PR64} to $\sim$
300\cite{Dyakonov_PRB92}, as compared with $\eta \sim 4000$ for GaAs.\cite{Paget_PRB77}

The study of hyperfine interaction in Si is currently drawing renewed interest because of the possibility of realizing spin qubits in Si.  For
example, in Ref.\onlinecite{Assali_preprint10}, an all-electron calculation was performed to obtain the complete hyperfine coupling matrix (both
contact and anisotropic parts).
We note that the pseudopotential method is not appropriate for calculating contact hyperfine interaction because it fails at the nuclear sites,
making the use of an all-electron calculation necessary. The calculated $\eta$ is about 160, consistent with experimental observations.
Furthermore, the calculated anisotropic hyperfine coupling strength is about 3\% of the contact part.  The finite contact hyperfine coupling
strength is a reflection that there is a finite component of atomic S-orbitals in the Bloch function at the bottoms of the Si conduction band,
even though the Bloch functions are predominantly P-like. The small calculated anisotropic hyperfine coupling strength is a reflection of the
highly symmetric nature of the Si lattice, in contrast with one-dimensional and two-dimensional structures such as carbon nanotubes and graphene
sheets. In short, hyperfine interaction in Si is an extensively studied subject, with its magnitudes in various situations already measured and
calculated.

An interesting contrast can be made between the hyperfine interactions in silicon and low-dimensional carbon structures, such as graphene. We
recall that the conduction band minima in Si are located at $\pm 0.85 \, (2\pi/a_{Si})$ along the (100), (010), and (001) axes. The six valleys
are in the vicinity of the X points, which are high-symmetry points, but not \textit{at} the X points themselves. Group theory tells us that the
Bloch functions near the X points are P-like. Similarly, the Bloch functions near the conduction band minima are predominantly P-like, but with
S-like and D-like admixtures.\cite{Koiller_PRB04} Nevertheless, group theory itself does not yield the weight of these admixtures, and numerical
calculations are required.

In contrast, the valleys in graphene are located at the high symmetry points ${\bm K}$ and ${\bm K}'$ in reciprocal space, where group theory
tells us the Bloch function are P-like.\cite{Dresselhaus_Group} When this is the case the contact hyperfine interaction vanishes. The dominant
hyperfine contribution arises from the anisotropic part of the interaction. This fact has interesting consequences for spin blockade, which were
explored in Refs. \onlinecite{Palyi_PRB09, Palyi_10}. Despite the many similarities between Si and C, since the valleys in Si are \textit{not}
located at high-symmetry points, symmetry places no additional constraints on the form of the wave functions. In short, unlike in graphene, the
contact hyperfine interaction in Si is finite,\cite{Assali_preprint10}
and it actually dominates over the anisotropic interaction.

For the present purpose, namely understanding the effect of the hyperfine interaction on spin and valley dynamics of singlet-triplet qubits in
Si quantum dots, it is sufficient to think of the hyperfine coupling as the interaction between an electron spin and an inhomogeneous Zeeman
field. Both the contact and the anisotropic terms can be cast in this form. This field will in general differ between adjacent quantum dots,
giving rise to singlet-triplet mixing when two states are close together in energy. The hyperfine interaction can in principle mix singlets and
triplets with the same valley composition as well as singlets and triplets with different valley composition. These cases will be examined in
what follows. We reiterate that any unwanted mixing present in natural Si can be eliminated by isotopic purification.

\subsubsection{Hyperfine-induced valley mixing}
\label{sec:hfvalley}

We write the contact hyperfine Hamiltonian in Si as
\begin{equation}
H_Z = - \mu_B \, [ {\bm \sigma}_1 \cdot {\bm B}_{nuc}({\bm r}_1) + {\bm \sigma}_2 \cdot {\bm B}_{nuc}({\bm r}_2)],
\end{equation}
where the hyperfine magnetic field can be written as
%
\begin{equation}\label{Bnuc}
{\bm B}_{nuc} =  \sum_i \, \frac{\mathcal{A}_{0}\nu_{0}}{\mu_B} \, \mathbf{I}_{i} \, \delta({\bm r} - {\bm R}_i).
\end{equation}
The contact interaction can mix states of different valley compositions. The dipolar hyperfine
(anisotropic) interaction for conduction electrons in Si is two orders of magnitude smaller,\cite{Assali_preprint10} so we neglect it for the following discussion.  We
again assume that a homogeneous magnetic field $\parallel \hat{\bm z}$ that is much larger than the nuclear field is applied so that we can
neglect mixing with polarized triplet states.  In other words, we only consider mixing of $\tilde{S}$ and $\tilde{T}^0$ here.

The intravalley matrix elements of the hyperfine interaction have the same form as those given above for the applied field.  We also have the
intervalley term
\begin{equation}
\begin{array}{rl}
\displaystyle \tbkt{D_z}{{\bm B}({\bm r})}{D_{\bar z}} = & \displaystyle {\bm b}^D.
\end{array}
\end{equation}
Since we are neglecting terms of the form $\tbkt{L}{\bm B}{R}$ and terms of linear and higher order in $g$ and $B$, $\tbkt{\tilde{D}}{\bm
B}{\tilde{D}} = \tbkt{D}{\bm B}{D}$ and we will write e.g. $\tbkt{\tilde{D}_+}{\bm B}{\tilde{D}_-} = {\bm B}^D_{+-}$:
\begin{equation}
\begin{array}{rl}
\displaystyle {\bm B}^D_{\pm\pm} = & \displaystyle {\bm B}_D \pm \frac{1}{2}\, (e^{i\phi_D}{\bm b}_D + e^{-i\phi_D}{\bm b}^*_D) \\ [3ex]
\displaystyle {\bm B}^D_{+-} = & \displaystyle - \frac{1}{2}\, (e^{i\phi_D}{\bm b}_D - e^{-i\phi_D}{\bm b}^*_D).
\end{array}
\end{equation}
In this work we take ${\bm B}^D$ and ${\bm b}^D$ as given.  These quantities can be determined by explicit evaluation of the matrix elements of
${\bm B}_{nuc}$ as given in Eq.\ (\ref{Bnuc}) using the wave functions $\{ D \}_\xi$ introduced in Sec.\ \ref{sec:DQD}, yielding
\begin{equation}\label{bd}
\begin{array}{rl}
\displaystyle {\bm B}_D = & \displaystyle \sum_{i, {\bm K}, {\bm Q}} \frac{\mathcal{A}_{0}\nu_{0} \mathbf{I}_{i}}{\mu_B} \, |F({\bm R}_i)|^2 \,
c^{z*}_{\bm K} c^z_{{\bm K} + {\bm Q}} \, e^{-i{\bm Q} \cdot{\bm R}_i} \\ [3ex] = & \displaystyle \sum_{i} \frac{\mathcal{A}\nu_{0}
\mathbf{I}_{i}}{\mu_B} \, |F({\bm R}_i)|^2  \\ [3ex] \displaystyle {\bm b}_D = & \displaystyle \sum_{i, {\bm K}, {\bm Q}}
\frac{\mathcal{A}_{0}\nu_{0} \mathbf{I}_{i}}{\mu_B} \, |F({\bm R}_i)|^2 \, c^{z*}_{\bm K} c^{\bar z}_{{\bm K} + {\bm Q}} \, e^{-i({\bm Q} +
2k_0\hat{\bm z}) \cdot{\bm R}_i}.
\end{array}
\end{equation}
Beyond this level the coefficients $c^\xi_{\bm K}$ in the expansions of the Bloch functions need to be determined.  Very recently $\mathcal{A}$
has been worked out in Ref.\ \onlinecite{Assali_preprint10}, which demonstrated that the contact hyperfine interaction is finite in a Si QD.
\cite{Assali_preprint10}. Explicit evaluation of 
${\bm b}^D$ requires more work. However, inspecting Eq.\ (\ref{bd}) we expect
most of the contribution to ${\bm B}_D$ to come from the term with ${\bm Q} = 0$. On the other hand in ${\bm b}_D$ the ${\bm Q} = 0$ term will
be much smaller, due to the modulation introduced by the fast oscillating factor $e^{-2ik_0Z_i}$. The sum over $i$ will bring this term very
close to zero, so that we expect ${\bm b}_D \ll {\bm B}_D$. We stress that, since we take these quantities as given (as indeed we do with
$\Delta$) we make no assumptions about the form of the Bloch functions in Si appearing in these matrix elements
(except for the assumption that the contact hf interaction is the dominant one).

We have the following interbranch matrix elements
\begin{equation}
\begin{array}{rl}
\displaystyle \tbkt{S^{LR}_{++}}{ H_{hf} }{T^{LR}_{+-}} = & \displaystyle \mu_B\, B^R_{z, +-} \\ [3ex]
\displaystyle \tbkt{S^{LR}_{--}}{ H_{hf} }{T^{LR}_{+-}} = & \displaystyle -\mu_B\, B^L_{z, -+} \\ [3ex]
\displaystyle \tbkt{S^{LR}_{++}}{ H_{hf} }{T^{LR}_{-+}} = & \displaystyle -\mu_B\, B^L_{z, +-} \\ [3ex]
\displaystyle \tbkt{S^{LR}_{--}}{ H_{hf} }{T^{LR}_{-+}} = & \displaystyle \mu_B\, B^R_{z, -+}  \\ [3ex]
\displaystyle \tbkt{S^{LR}_{++}}{ H_{hf} }{T^{LR}_{sym}} = & \displaystyle -\mu_B\, \frac{\Delta B_{z, +-}}{\sqrt{2}} \\ [3ex]
\displaystyle \tbkt{S^{LR}_{++}}{ H_{hf} }{T^{LR}_{anti}} = & \displaystyle \mu_B\, \frac{B^{tot}_{z, +-}}{\sqrt{2}} \\ [3ex]
\displaystyle \tbkt{S^{LR}_{--}}{ H_{hf} }{T^{LR}_{sym}} = & \displaystyle -\mu_B\, \frac{\Delta B_{z, -+}}{\sqrt{2}} \\ [3ex]
\displaystyle \tbkt{S^{LR}_{--}}{ H_{hf} }{T^{LR}_{anti}} = & \displaystyle -\mu_B\, \frac{B^{tot}_{z, -+}}{\sqrt{2}}.
\end{array}
\end{equation}
In the matrix elements above the singlet and triplet labels could be reversed and the value will remain the same.  Doing this for the first
matrix element we obtain $\tbkt{T^{LR}_{++}}{ H_{hf} }{S^{LR}_{+-}} = \mu_B\, B^R_{z, +-}$, and likewise for the subsequent ones. The hyperfine
interaction does not mix the $++$ and $--$ branches directly. Within the $+-$ branch we have the intervalley matrix elements
\begin{equation}
\begin{array}{rl}
\displaystyle \tbkt{S^{LR}_{+-}}{ H_{hf} }{T^{LR}_{+-}} = & \displaystyle B^L_{z, ++} - B^R_{z, --} \\ [3ex]
\displaystyle \tbkt{S^{LR}_{-+}}{ H_{hf} }{T^{LR}_{-+}} = & \displaystyle B^L_{z, --} - B^R_{z, ++} \\ [3ex]
\displaystyle \tbkt{S^{LR}_{+-}}{ H_{hf} }{T^{LR}_{-+}} = & \displaystyle \tbkt{\psi^{LR}_{S-+}}{ H_{hf} }{\psi^{LR}_{T_0+-}} = 0 \\ [3ex]
\displaystyle \tbkt{S^{LR}_{sym}}{ H_{hf} }{T^{LR}_{sym}} = & \displaystyle \frac{1}{2}\, (\Delta B_{z, ++} + \Delta B_{z, --}) \\ [3ex]
\displaystyle \tbkt{S^{LR}_{anti}}{ H_{hf} }{T^{LR}_{anti}} = & \displaystyle \frac{1}{2}\, (\Delta B_{z, ++} + \Delta B_{z, --}) \\ [3ex]
\displaystyle \tbkt{S^{LR}_{sym}}{ H_{hf} }{T^{LR}_{anti}} = & \displaystyle \frac{1}{2}\, (B^{tot}_{z, ++} - B^{tot}_{z, --}).
\end{array}
\end{equation}
The states ${T^{LR}_{sym/anti}}$ and ${S^{LR}_{sym/anti}}$ are defined in the following Section.  Unless specified otherwise, all remaining
matrix elements are zero.

\subsection{Spin and valley blockade in multi-valley Si QDs}

Spin blockade allows charge sensors (quantum point contact or single-electron transistor) to differentiate the two-electron singlet and triplet
states.\cite{Ono_Science02, Hanson_RMP07, Petta_Science05} In this subsection we explain the details of state mixing in the $++/--/+-$ branches,
and the reason an inhomogeneous magnetic field leads to spin blockade of only certain states. We also consider spin blockade in a hyperfine
field, which has matrix elements between different valleys, and study their effect on singlet-triplet qubit operation.

Spin blockade occurs when the electrons occupy the (1,1) state with spin/valley structure incompatible with the structure of available (0,2)
states.  In such a case upon changing the detuning from the (1,1) regime to the nominally (0,2) regime tunneling of electrons into the right dot
is suppressed, and the charge state of the DQD remains as (1,1). In the single-valley case this happens when we have a (1,1) triplet, which is
lower in energy compared to the (0,2) triplet state after changing the interdot detuning to the detection regime, where the (0,2) singlet is the
ground state. Therefore an initial (1,1) singlet state evolves to (0,2), while the (1,1) triplet state stays in the (1,1) configuration. This
difference in the charge distribution can be detected by a charge sensor. In the multi-valley case the situation is more complicated.

We first consider an applied inhomogeneous magnetic field, and study the $++/--$ branches, which is described by the Hamiltonian $H_{\pm\pm}$.
For $\delta \ll \delta_c$ the eigenstates are $\tilde{S}^{RR}_{\pm\pm}$, $\tilde{S}^{LR}_{\pm\pm}$ and $\tilde{T}^{LR}_{\pm\pm}$.  As the
detuning is swept adiabatically from $\delta \ll \delta_c$ to $\delta \gg \delta_c$, the singlets $\tilde{S}^{LR}_{\pm\pm}$ evolve into
$\tilde{S}^{RR}_{\pm\pm}$, while the triplets $\tilde{T}^{LR}_{\pm\pm}$ remain unchanged: the (1,1) singlet evolves into the (0,2) singlet,
while the (1,1) triplet remains in the (1,1) configuration, as in GaAs.

In the $+-$ branch, spin blockade is rather subtle and interesting.  Consider the singlet branch [$H^S_{+-}$ of Eq.~(\ref{eq:HS+-}).]  The
eigenstates in the (0,2) and the far-detuned (1,1) regimes  are respectively $\tilde{S}^{RR}_{+-}$ and
\begin{equation}
\begin{array}{rl}
\displaystyle \tilde{S}^{LR}_{sym} = & \displaystyle \frac{1}{\sqrt{2}} \, \big( \tilde{S}^{LR}_{+-} + \tilde{S}^{LR}_{-+} \big) \\ [2ex]
\displaystyle \tilde{S}^{LR}_{anti} = & \displaystyle \frac{1}{\sqrt{2}} \, \big( \tilde{S}^{LR}_{+-} - \tilde{S}^{LR}_{-+} \big).
\end{array}
\end{equation}
In the regime of $\delta \sim \delta_c$, only $\tilde{S}^{RR}_{+-}$ and $\tilde{S}^{LR}_{sym}$ are tunnel-coupled. Therefore, as the detuning is
swept from $\delta - \delta_c \ll 0$ to $\delta - \delta_c \gg 0$, the ground state evolves from $\tilde{S}^{LR}_{sym}$ into
$\tilde{S}^{RR}_{+-}$, while $\tilde{S}^{LR}_{anti}$ remains unchanged. For the $+-$ triplet branch [$H^T_{+-}$ of Eq.~(\ref{eq:HS+-})], the
eigenstates in the far-detuned regimes of $\delta \ll \delta_c$ and $\delta \gg \delta_c$ are $\tilde{T}^{RR}_{+-}$ and
\begin{equation}
\begin{array}{rl}
\displaystyle \tilde{T}^{LR}_{sym} = & \displaystyle \frac{1}{\sqrt{2}} \, \big( \tilde{T}^{LR}_{+-} + \tilde{T}^{LR}_{-+} \big) \\ [2ex]
\displaystyle \tilde{T}^{LR}_{anti} = & \displaystyle \frac{1}{\sqrt{2}} \, \big( \tilde{T}^{LR}_{+-} - \tilde{T}^{LR}_{-+} \big).
\end{array}
\end{equation}
As the detuning is swept from $\delta \ll \delta_c$ to $\delta \gg \delta_c$, the anti-symmetric state $\tilde{T}^{LR}_{anti}$ evolves into
$\tilde{T}^{RR}_{+-}$ while $\tilde{T}^{LR}_{sym}$ remains unchanged.  This is the opposite of what happens in the $+-$ singlet branch.

Operation of the singlet-triplet qubit entails initialization of a unique (0, 2) state.  When $\tilde{S}^{RR}_{+-}$ is initialized and the
detuning is swept from $\delta \gg \delta_c$ to $\delta \ll \delta_c$, this state evolves into $\tilde{S}^{LR}_{sym}$. Under the action of an
applied inhomogeneous magnetic field this state can only mix with the triplet $\tilde{T}^{LR}_{sym}$. On the other hand, when
$\tilde{T}^{RR}_{+-}$ is initialized and the detuning is swept from $\delta \gg \delta_c$ to $\delta \ll \delta_c$, this state evolves into
$\tilde{T}^{LR}_{anti}$, which, in an inhomogeneous magnetic field, can only mix with the triplet $\tilde{S}^{LR}_{anti}$. It is the same
component of the inhomogeneous magnetic field that mixes $\tilde{S}^{LR}_{sym}$ with $\tilde{T}^{LR}_{sym}$ and $\tilde{T}^{LR}_{anti}$ with
$\tilde{S}^{LR}_{anti}$. We may refer to this process as \textit{valley blockade}, similar to that seen in carbon nanotube
QDs.\cite{Palyi_PRB09, Palyi_10} As we sweep the detuning and allow states to mix, the destination (1,1) states ($\tilde{T}^{LR}_{sym}$ and
$\tilde{S}^{LR}_{anti}$) do not tunnel couple back to the (0,2) configuration, so that they are blocked. This should then allow a charge sensor
measurement of the system states.

Consider now an inhomogeneous nuclear field. To lowest order, the $++$ and $--$ branches do not mix, yet they can both mix with the $+-$ branch.
Some states cross, therefore they can be close enough in energy to be mixed. However the time scale for this process is on the order of 0.1-10
$\mu$s,\cite{Assali_preprint10} whereas experiment uses rapid adiabatic passage, in which the system generally remains in the (1,1) regime for
up to a few tens of ns, thus mixing of states between different branches will be avoided at the points where the $--$ and $+-$ levels cross (see
Fig~\ref{LargeDelta_All}.) At $\delta \ll \delta_c$, for mixing between different branches to be avoided, we need $|\Delta| \gg |B_{nuc}|$,
which is satisfied by any $\Delta$ an order of magnitude larger than a few neV.

Within the $+-$ branch, consider initialization of a $(0,2)$ state.  The singlet evolves into $\tilde{S}_{sym}$ and the triplet evolves into
$\tilde{T}_{anti}$. These two states do not mix in an applied inhomogeneous field, yet they can mix under the influence of the Overhauser field.
If for example $\tilde{S}^{RR}_{+-}$ is initialized and evolves into $\tilde{S}^{LR}_{sym}$, in the Overhauser field this state mixes with
$\tilde{T}^{LR}_{sym}$ (matrix element ${\bm B}_D$) or $\tilde{T}^{LR}_{anti}$ (matrix element ${\bm b}_D$.) If both processes are of the same
order, the former remains in (1,1) but the latter goes back to (0,2) -- it is neither spin blocked nor valley blocked, and readout of the charge
state of the right dot would not give any indication of whether mixing had occurred in the (1,1) configuration (yet if one starts out with
$\tilde{S}^{RR}_{+-}$ one can never end up with $\tilde{S}^{LR}_{anti}$.) We have seen that ${\bm b}_D$ is expected to be much smaller than
${\bm B}_D$, therefore the mixing of $\tilde{S}^{LR}_{sym}$ and $\tilde{T}^{LR}_{anti}$ will take place on time scales much longer than 1$\mu$s.
Nevertheless, to determine the probabilities of return for natural Si one requires the intervalley matrix elements ${\bm b}^D$ of the hyperfine
interaction. More work is needed to determine these more quantitatively. However, the $^{29}$Si can be isotopically purified, reducing the
hyperfine-induced valley mixing as much as is needed. Under these circumstances we expect valley mixing to have little or no effect on spin
blockade.

We note that spin blockade is related to the so-called supertriplet and supersinglet states in the context of graphene quantum
dots.\cite{Palyi_PRB09} In the context of Si quantum dots studied in this work, there are 16 states in the (1,1) configuration.  Inspection of
Fig.~\ref{LargeDelta_All} shows that, when the detuning is swept from $\delta \ll \delta_c$ to $\delta \gg \delta_c$, 10 of these states do not
change to the (0,2) configuration: the nine triplets $\tilde{T}^{LR}_{++}$, $\tilde{T}^{LR}_{--}$, and $\tilde{T}^{LR}_{sym}$ as well as the
singlet $\tilde{S}^{LR}_{anti}$. These are the 10 \textit{supertriplet} states identified in Ref.\ \onlinecite{Palyi_PRB09}. The 6
\textit{supersinglet} states, which do switch to the (0,2) configuration, are the singlets $\tilde{S}^{LR}_{++}$, $\tilde{S}^{LR}_{--}$, and
$\tilde{S}^{LR}_{sym}$, plus the three triplets $\tilde{T}^{LR}_{anti}$. Similar conclusions regarding spin blockade in carbon nanotube quantum
dots were reached in Refs.\ \onlinecite{Palyi_PRB09, Palyi_10}.

\subsection{Single-qubit gates}

For $|\tilde{\Delta}| \gg k_BT$ singlet-triplet qubits in Si can be operated in the same way as in GaAs.  The physics becomes more interesting
when the valley splitting is of the order of $k_BT$ (though this case makes quantum information processing less practical), so that any one of
six initial states may be loaded. As the detuning is varied the system may go through any of the three anticrossings in
Fig.~\ref{LargeDelta_All}. If any of the singlet or unpolarized triplet states is initialized, the outcome of sweeping the detuning will be the
same as in GaAs. The polarized triplets $\tilde{T}^{\pm,RR}_{+-}$ are split off by a uniform Zeeman field that exceeds the inhomogeneous Zeeman
field by a large amount by construction. The time scale for these states to be mixed with any other states by the inhomogeneous Zeeman field is
thus extremely long, and much longer than the time of the experiment. As a result these states will always have a probability of return of 1.
Overall, it is as if four independent qubit channels existed in a DQD, and for small $|\tilde{\Delta}|$ one could load any of them (with the
possible exception of the higher-energy polarized triplet states). An experiment analogous to Ref.~\onlinecite{Petta_Science05} is feasible,
except in Si it will be carried out on a series of independent qubit channels.

Quantum computation requires both $\sigma_x$ and $\sigma_z$ gates to achieve arbitrary one-qubit rotations.  The inhomogeneous magnetic field
provides the $\sigma_x$ gate, which has been discussed, while $\sigma_z$ gates are feasible by means of exchange. The case $|\tilde{\Delta}| \gg
k_BT$ is analogous to GaAs and $\sigma_z$ gates can be implemented as in Ref.\ \onlinecite{Petta_Science05}. Let us focus on $\sigma_z$ gates
when $\tilde{\Delta} = 0$. Once the polarized triplet states have been separated, four states remain (effectively) degenerate:
$\tilde{S}^{RR}_{\pm\pm}$, $\tilde{S}^{RR}_{+-}$ and $\tilde{T}^{RR}_{+-}$.  If $\tilde{S}^{RR}_{\pm\pm}$ are loaded, in the far detuned regime
they will be separated from $\tilde{T}^{LR}_{\pm\pm}$ by $\tilde{J}$. If one initializes into the $+-$ branch one will have a superposition of
$\tilde{S}^{RR}_{+-}$ and $\tilde{T}^{RR}_{+-}$. In the far detuned regime these states evolve into $\tilde{S}_{sym}$ and $\tilde{T}_{anti}$.
Nearest in energy to these states we have $\tilde{S}_{anti}$ and $\tilde{T}_{sym}$, which are also degenerate, and split from $\tilde{S}_{sym}$
and $\tilde{T}_{anti}$ by $\tilde{J}$. In other words, there is only one exchange $\tilde{J}$. In general, one can initialize into a
superposition of the four states above. Yet each state mixes with only one other state, and the energy splitting between each pair of states is
$\tilde{J}$. Thus, no matter what superposition of states is loaded, exchange-based $\sigma_z$ gates can be obtained.

It was demonstrated in Sec.~\ref{sec:Loss_PRA98} that exchange-based gates in single-spin qubits are not feasible unless $|\tilde{\Delta}| \gg
k_BT$, whereas here they are feasible for any $|\tilde{\Delta}|$. In the LDV architecture the qubits are initialized independently of each
other. One may prepare both qubits in the same state ($\tilde{L}_+$ and $\tilde{R}_+$ or $\tilde{L}_-$ and $\tilde{R}_-$), then allow the
exchange gate to rotate the corresponding singlets and triplets ($\tilde{S}^{LR}_{\pm\pm}$ and $\tilde{T}^{LR}_{\pm\pm}$) into each other. If
the two qubits are prepared in different branches we have either the singlets and triplets $\tilde{S}^{LR}_{+-}$ and $\tilde{T}^{LR}_{+-}$, or
$\tilde{S}^{LR}_{-+}$ and $\tilde{T}^{LR}_{-+}$. We have seen above that these states are degenerate: exchange splitting between is always zero.
On the other hand, in singlet-triplet qubits, in the (1,1) configuration one always accesses the states $\tilde{S}_{sym}$, $\tilde{T}_{sym}$,
$\tilde{S}_{anti}$ and $\tilde{T}_{anti}$. These are superpositions of $\tilde{S}^{LR}_{+-}$ and $\tilde{S}^{LR}_{-+}$, and
$\tilde{T}^{LR}_{+-}$ and $\tilde{T}^{LR}_{-+}$. The cross terms between the $+-$ and $-+$ states ensure that exchange between
$\tilde{S}_{sym}/\tilde{T}_{sym}$ and between $\tilde{S}_{anti}/\tilde{T}_{anti}$ is always $\tilde{J}$.

\subsection{Two-qubit operations and scalability}

One proposal for two-qubit operations has been described in Ref.\ \onlinecite{Taylor_NP05}, closely related to the exchange-based scheme of
Ref.\ \onlinecite{Levy_PRL02}. It makes use of the fact that the singlet ground state is a superposition of the (1,1) and (0,2) configurations,
while the triplet ground state within the operational/measurement regime is in the (1,1) configuration. The difference in charge distribution
allows state-sensitive coupling. Coulomb interaction between electrons on adjacent DQDs results in an effective dipolar interaction that
entangles qubits. In a one-valley system, for two qubits A and B we may construct the two-qubit states $\tilde{S}_A\tilde{S}_B$,
$\tilde{T}_A\tilde{T}_B$, $\tilde{T}_A\tilde{S}_B$, and $\tilde{S}_A\tilde{T}_B$. If electrons 1 and 2 are in qubit $A$ while 3 and 4 are in
qubit $B$, the matrix elements of the Coulomb interaction that lead to entanglement are
\begin{equation}
\begin{array}{rl}
\displaystyle V_{SS} = & \displaystyle \tbkt{\tilde{S}_A\tilde{S}_B}{V^{AB}_{ee}}{\tilde{S}_A\tilde{S}_B} \\ [3ex]
\displaystyle V_{TT} = & \displaystyle \tbkt{\tilde{T}_A\tilde{T}_B}{V^{AB}_{ee}}{\tilde{T}_A\tilde{T}_B} \\ [3ex]
\displaystyle V_{TS} = & \displaystyle \tbkt{\tilde{T}_A\tilde{S}_B}{V^{AB}_{ee}}{\tilde{T}_A\tilde{S}_B} \\ [3ex]
\displaystyle V_{ST} = & \displaystyle \tbkt{\tilde{S}_A\tilde{T}_B}{V^{AB}_{ee}}{\tilde{S}_A\tilde{T}_B} \\ [3ex]
\displaystyle V^{AB}_{ee} \equiv & \displaystyle V_{ee}(1,3) + V_{ee}(1,4) + V_{ee}(2,3) + V_{ee}(2,4),
\end{array}
\end{equation}
with $V_{ee}(i,j)$ the Coulomb interaction between electrons $i$ and $j$.  We use the generic notation $V_{XY}$ for these terms, where $X, Y =
S, T$.

In Si the DQD spectrum consists of three branches. If $|\tilde{\Delta}| \le k_BT$ and ambiguity exists in initialization, then states from all
branches may be loaded. In Si, if both qubits belong to the same branch, one encounters the terms
\begin{equation}
\begin{array}{rl}
\displaystyle V_{XY} (\pm\pm, \pm\pm) = & \displaystyle \tbkt{\tilde{X}^{\pm\pm}_A\tilde{Y}^{\pm\pm}_B}{V^{AB}_{ee}}{\tilde{X}^{\pm
\pm}_A\tilde{Y}^{\pm\pm}_B} \\ [3ex] \displaystyle V_{XY} (+-, +-) = & \displaystyle
\tbkt{\tilde{X}^{+-}_A\tilde{Y}^{+-}_B}{V^{AB}_{ee}}{\tilde{X}^{+-}_A\tilde{Y}^{+-}_B}.
\end{array}
\end{equation}
Since the Coulomb interaction has no matrix elements between $z$ and $\bar{z}$ states $V_{XY} (++,++) = V_{XY} (--,--) = V_{XY} (+-, +-)$.  It
does not make a difference which branch is loaded as long as both qubits are loaded in the same branch. At the same time, if the qubits are
loaded in different branches, the interaction between them will be governed by matrix elements of the form
\begin{equation}
\begin{array}{rl}
\displaystyle V_{XY} (\pm\pm, \mp\mp) = & \displaystyle \tbkt{\tilde{X}^{\pm\pm}_A\tilde{Y}^{\mp\mp}_B}{V^{AB}_{ee}}{\tilde{X}^{\pm
\pm}_A\tilde{Y}^{\mp\mp}_B} \\ [3ex] \displaystyle V_{XY} (+-, -+) = & \displaystyle
\tbkt{\tilde{X}^{+-}_A\tilde{Y}^{-+}_B}{V^{AB}_{ee}}{\tilde{X}^{+-}_A\tilde{Y}^{-+}_B},
\end{array}
\end{equation}
which will all be suppressed by exponential prefactors as we have discussed, and will be effectively zero.  The matrix elements of the Coulomb
interaction between DQDs will be the same for qubits belonging to the same branch, but will be zero for qubits belonging to different branches.
When $|\tilde{\Delta}|$ is comparable to $k_BT$ one can no longer be certain of initializing fiducially into a particular state. The matrix
element of the Coulomb interaction depends on the explicit superposition loaded. If one loads qubit $A$ as $\alpha_A\tilde{S}_A^{++} +
\beta_A\tilde{S}_A^{--} + \gamma_A \tilde{S}_A^{+-} + \delta_A \tilde{T}_A^{0,+-}$ and qubit $B$ as $\alpha_B\tilde{S}_B^{++} +
\beta_B\tilde{S}_B^{--} + \gamma_B \tilde{S}_B^{+-} + \delta_B \tilde{T}_B^{0,+-}$ into the second, the matrix elements $V_{XY}$ depend on the
explicit values of the coefficient $\alpha_{A, B}$, $\beta_{A,B}$, $\gamma_{A,B}$, $\delta_{A,B}$. Coulomb interaction-based entanglement in
singlet-triplet qubits suffers from the same problem as single-spin qubits: it will work if $|\tilde{\Delta}| \gg k_BT$, but will fail for
$|\tilde{\Delta}| \le k_BT$.

While single-qubit operations are formally still possible even with the multiplicity introduced by the valleys, it does inhibit two-qubit
operations and therefore quantum computation. The fundamental reason is that the Coulomb interaction between states belonging to different
branches of the spectrum is vastly different. Therefore, for single-spin and singlet-triplet quantum computers to work, it is imperative that
$|\tilde{\Delta}|$ be much greater than $k_BT$. We mention for completeness that the valley degree of freedom may not be a problem if two-qubit
operations are implemented by other means, not relying on exchange, such as a Jaynes-Cumings type interaction familiar from atomic physics, yet
more work is needed to establish a quantitative analysis of such schemes in Si.

\section{Interface roughness}
\label{sec:position}

In general $\tilde{\Delta}$ is complex \cite{Saraiva_PRB09, Hada_JJAP04} and is characterized by an amplitude and a phase.  Interface roughness
will introduce into the problem an additional random potential. Such a random potential would yield a correction to $\tilde{\Delta}$ and could
make the amplitude and phase of $\tilde{\Delta}$ different in the two dots. This difference in valley composition would enable intervalley
tunneling during interdot transitions, which could in principle affect two-dot dynamics, in particular the operations of a singlet-triplet
qubit. Problems could also arise if the exchange coupling is different in the three different branches $++$, $--$ and $+-$. Such a difference
could be induced by interface roughness. We will address these issues elsewhere.\cite{Culcer_Roughness_10} Meanwhile, we would like to comment that interface
roughness is a physical quantity that is difficult to determine \textit{a priori} without destroying the device. Still, it is something that one
has to keep in mind when considering spin physics in Si quantum dots.

\section{Other types of qubits}
\label{sec:other}

Given the additional valley degree of freedom present in silicon, it is natural to ask whether silicon quantum dots offer multiple possibilities
for implementing and initializing qubits. We discuss two other possibilities different from the situation considered thus far, and show that,
based on our current understanding and on the current ability to manipulate Si QDs in the laboratory, these alternatives cannot produce a
practical qubit.

Firstly, it is evident from our discussion that the only way one can be certain of initializing a particular state in the $(0,2)$ configuration
without prior knowledge of valley splitting is to apply a strong homogeneous magnetic field so as to push the triplet $\tilde{T}^{+, RR}_{+-}$
below all the other states by an amount greatly exceeding $k_B T$. The detuning can then be altered and one electron will tunnel to form the
state $\tilde{T}^{+, LR}_{+-}$. In order to mix this state with another state it will be necessary to reduce the magnetic field greatly so that
this state is close enough in energy to another state and the inhomogeneous magnetic field can mix them. In addition, the three triplet states can
only be mixed by a uniform transverse magnetic field, but not by an inhomogeneous field. In the end we have a 4-level system that has
complicated and sometimes hard to control couplings, making it impractical for quantum information processing applications.

One could also consider using the valley degree of freedom to construct a qubit, in other words to employ the two valley-split states as a two
level system and attempt to perform operations on it. In this case the same initialization problems described above are present. In addition,
there is practically no control over the valley splitting and correspondingly no control over single qubits: there is no way to rotate one qubit
coherently and controllably once it is initialized. The interesting paradox in this situation is that one could, in principle, implement
two-qubit operations by using the exchange interaction provided by the valley-exchange Coulomb integral. This valley-exchange Coulomb integral
has been discussed above and could be used to rotate states with different valley character into each other. Nevertheless, given the little
amount of control over the valley splitting, and that neither theory nor experiment have been able to determine the exact size of the
valley-exchange Coulomb integral $\tilde{j}_v$, this would be a highly impractical method at present.

\section{Identifying valley-split states}  \label{sec:identyfying}

To implement a qubit in a reliable manner we need to determine whether the condition $|\tilde{\Delta}| \gg k_B T$ is satisfied.  We recall that
$\tilde{\Delta}$ has been calculated recently \cite{Saraiva_PRB09} and it is believed that its magnitude can be as high as $\approx$0.25 meV,
and can be further increased by applying an electric field. In Ref. \onlinecite{Culcer_PRB09} we described a quantum coherent experiment to
measure the valley splitting. An alternative approach to measure $|\tilde{\Delta}|$ is to use a two-electron single quantum dot and probe its
ground state while sweeping the magnetic field. The basic idea here again relies on the detection of the magnetic field at which the ground
state undergoes a singlet-triplet transition. This single-dot experiment requires charge sensing.  One should adjust the gate potential in such
a way that the dot contains two electrons, so that we can refer once more to the spectrum of the doubly-occupied single dot as in Fig.\
\ref{SingleDotLevels}. One can then perform tunnelling spectroscopy to determine the spectrum of the QD, and identify the ground state.  At zero
applied field the ground state should be a $\tilde{S}_{--}$ singlet. As the applied field is increased, the ground state remains to be the
singlet and barely changes in energy with the field, until it switches to $\tilde{T}^{+,RR}_{+-}$, beyond which the ground state energy energy
would decrease linearly with the increasing field. This singlet-triplet transition in the ground state occurs when $E_Z = 2|\tilde{\Delta}|$.
Recording the ${\bm B}$-field dependence of the ground state energy and identifying ${\bm B}$ for the singlet-triplet transition gives a
reliable estimate of $|\tilde{\Delta}|$ in this QD. This description also shows that the experiment can be reformulated alternatively as a
transport (resonant tunnelling) experiment. Recent experimental work on Si QDs \cite{Lim_APL09, Lim_SingleElectron_APL09} suggests that this
experiment is feasible at present. We are indebted to discussions with Malcolm Carroll for this idea.

\section{Summary and Conclusions}

We have analyzed the way the valley degree of freedom affects spin qubits in Si quantum dots. The multiplicity of the ground state alters the
energy level spectrum that is accessed in the implementation of single-spin and singlet-triplet qubits. Moreover, unlike in single-valley
systems, the energy scale is no longer set by the confinement energy, but by the valley splitting, which is expected to be approximately one
order of magnitude lower.

Initialization is a crucial step for any qubit. Fo Si QDs, when $|\tilde{\Delta}| \gg k_BT$, the system is effectively a one-valley system, and
both single-spin and singlet-triplet qubits are analogous to those in GaAs.

When the valley splitting $|\tilde{\Delta}|$ is on the order of $k_B T$ or smaller, both single-spin and two-spin qubit architectures become
essentially impossible. In particular, for a single-spin qubit, different single-electron valley eigenstates cannot be energetically
distinguished, so that reliable orbital initialization cannot be realized even if spin states can be properly prepared. While single-spin
rotations can still be performed, the existence of non-identical orbital states will make the exchange coupling between neighboring quantum dots
unpredictable and dramatically increase the difficulty of two-qubit gates. Therefore spin quantum computation becomes practically impossible in
this limit.

In singlet-triplet qubits, if $|\tilde{\Delta}| \le k_BT$ and provided a uniform Zeeman field separates the polarized triplet states, any one of
four two-electron states may be loaded on a single dot: two singlet states in which the electrons are in the same valley eigenstate, as well as
one singlet and one unpolarized triplet state in which the electrons are in different valley eigenstates. The DQD spectrum is characterized by
three distinct anticrossings, which represent three different branches of the spectrum, each with a different valley composition. Depending on
the relative values of $|\tilde{\Delta}|$, the Zeeman energy $E_Z$, and tunnel coupling $\tilde{t}$, these branches may interpenetrate,
resulting in a complex experimental energy level spectrum. An applied magnetic field does not mix valley eigenstates with different valley
composition. In a nuclear magnetic field the dominant term is due to the contact hyperfine interaction. An intervalley term exists that in
principle lifts the spin blockade in the middle (valley-mixing) branch of the spectrum. We expect this term to be much smaller than the
intravalley matrix element of the hyperfine interaction, which is the relevant matrix element for $\sigma_x$ gates. Therefore, although singlets
and triplets from different branches of the spectrum may mix in an Overhauser field, this mixing occurs on time scales much longer than those
relevant to quantum coherent experiments and $\sigma_x$ gates are feasible. Furthermore, since exchange is the same in all three branches of the
DQD spectrum, $\sigma_z$ gates can also be implemented even if $|\tilde{\Delta}| \le k_BT$. For $|\tilde{\Delta}|$ comparable to $k_BT$ the
system could be viewed as a set of four independent qubit channels. At the same time, the strategy for scaling up singlet-triplet qubits
proposed in Ref.\ \onlinecite{Taylor_NP05} is not practicable when $|\tilde{\Delta}| \le k_BT$, since the Coulomb interaction is again
unpredictable between neighboring qubits. Yet spin blockade in Si in an applied inhomogeneous magnetic field occurs in the same way as in GaAs.

For both single-spin and singlet-triplet qubits, one-qubit operations are feasible in the presence of a valley degree of freedom.  Two-qubit
operations based on the Coulomb interaction are only possible if both qubits have the same valley composition, and depend on the ability to
initialize all qubits unambiguously into the same state. Architectures in which entanglement relies on on the Coulomb interaction are not
scalable unless $|\tilde{\Delta}| \gg k_BT$. For these schemes the only feasible option is to ensure $|\tilde{\Delta}| \gg k_BT$ and implement
spin qubits in the same way as in a single-valley system such as GaAs.

We have shown also that other types of qubits involving the valley degree of freedom are not feasible based on our current knowledge and ability
to control Si QDs in the laboratory.

Valley-split states and the magnitude of the valley splitting may be identified by sweeping a magnetic field. This can be done using charge
sensing or transport of two electrons on a single quantum dot, or by performing a quantum coherent experiment on a singlet-triplet qubit in a
double quantum dot.\cite{Culcer_PRB09} These measurements can be checked against one another.

The valley degree of freedom of Si poses different qualitative problems for Si QD spin qubits compared with the corresponding Si:P bulk spin
qubits, where valley degeneracy leads to the exchange oscillation phenomenon, making the fabrication and control of two-qubit exchange gates
problematic. \cite{Koiller_PRL01, Koiller_PRB02} In that case, it is the lifting of the valley degeneracy (rather than the valley degeneracy
itself) by the singular Coulomb potential of the donors that is at fault. It appears that the valley degree of freedom is a serious enemy of
solid state spin quantum computation independent of whether one considers surface or bulk spin qubits.  In fact, other solid state systems where
valley degeneracy exists, such as graphene or carbon nanotubes, may also have difficulties with respect to spin qubit architecture because of
the competition between spin and valley degrees of freedom.

For a few qubits the existence of valleys closely spaced in energy is not necessarily a problem, since experiment can use trial and error and
explicitly characterize each sample, ensuring that the valley splitting is sufficiently large. Nevertheless the quantum computer architectures
discussed in this work are not scalable unless one determines a reliable way to control the valley splitting systematically. The ideal solution
would be to develop a quantitatively precise control over valley splitting so that it can routinely be engineered to be large. Unfortunately,
designing samples with large valley splitting thus remains an elusive task.

\begin{acknowledgements}

We thank M.~S.~Carroll for suggesting the experiment for measuring the valley splitting. We acknowledge enlightening discussions with
N.~M.~Zimmerman, J.~M.~Taylor, Ted Thorbeck, Jason Kestner, M.P. Lilly, Erik Nielsen, Lisa Tracy, Malcolm Carroll, H.~W.~Jiang, Matt House, Mark
Eriksson, Mark Friesen, S.N. Coppersmith, C. B. Simmons, A.T. Hunter, R. Joynt, Q. Niu and Zhenyu Zhang. This work is supported by LPS-NSA-CMTC.
{\L}C also acknowledges support from the Homing programme of the Foundation for Polish Science supported by the EEA Financial Mechanism. XH
acknowledges support by NSA/LPS through ARO.

\end{acknowledgements}

\appendix

\begin{widetext}

\section{Matrix elements}\label{app:matrixel}

The interdot tunneling matrix element $t$ has a single-particle part $t_0$ and a Coulomb-interaction (enhancement) part $s$, and we write it as
$t = t_0 + s$. Starting with the wave functions $\ket{D_\xi}$ for a single electron in a Si DQD, the matrix elements of the Hamiltonian within
the Hund-Mulliken approximation involve the following integrals (the superscript ${(i)}$ labels the $i$-th electron)
\begin{equation}
\arraycolsep 0.3 ex
\begin{array}{rl}
\displaystyle t_0 = & \displaystyle \tbkt{L_\xi}{H_0}{R_\xi} \\ [1ex] \displaystyle k = & \displaystyle \tbkt{L_\xi^{(1)}
R_\xi^{(2)}}{V_{ee}}{L_\xi^{(1)}  R_\xi^{(2)}} = \tbkt{L_\xi^{(1)} R_{-\xi}^{(2)}}{V_{ee}}{L_\xi^{(1)}  R_{-\xi}^{(2)}}  \\ [1ex] \displaystyle
j = & \displaystyle \tbkt{L_\xi^{(1)}  R_\xi^{(2)}}{V_{ee}}{L_\xi^{(2)}  R_\xi^{(1)}} = \tbkt{L_\xi^{(1)}
R_{-\xi}^{(2)}}{V_{ee}}{L_{-\xi}^{(2)}  R_\xi^{(1)}} \\ [1ex] \displaystyle s = & \displaystyle \tbkt{L_\xi^{(1)}
L_\xi^{(2)}}{V_{ee}}{L_\xi^{(1)}  R_\xi^{(2)}} = \tbkt{L_\xi^{(1)}  L_{-\xi}^{(2)}}{V_{ee}}{L_\xi^{(1)}  R_{-\xi}^{(2)}} \\ [1ex] \displaystyle
u = & \displaystyle \tbkt{D_\xi^{(1)}  D_\xi^{(2)}}{V_{ee}}{D_\xi^{(1)}  D_\xi^{(2)}} = \tbkt{D_\xi^{(1)}  D_{-\xi}^{(2)}}{V_{ee}}{D_\xi^{(1)}
D_{-\xi}^{(2)}}.
\end{array}
\end{equation}
We shall refer to these as the bare matrix elements. On an asymmetric double dot we have to allow for the energies on the left and right dots,
$\varepsilon_L$ and $\varepsilon_R$, to be different, as explained in the text. When one switches from $\{ \ket{L}, \ket{R} \}$ to the
orthogonal basis spanned by $\{ \ket{\tilde{L}}, \ket{\tilde{R}} \}$, the matrix elements of the Hamiltonian involve the following integrals
\begin{equation}
\arraycolsep 0.3 ex
\begin{array}{rl}
\displaystyle \tilde{\varepsilon}_L = & \displaystyle \frac{\varepsilon_L + g^2\varepsilon_R - 2gt_0}{1 - 2lg + g^2} \\ [3ex]
\displaystyle \tilde{\varepsilon}_R = & \displaystyle \frac{\varepsilon_R + g^2\varepsilon_L - 2gt_0}{1 - 2lg + g^2} \\ [3ex]
\displaystyle \tilde{t}_0 = & \displaystyle \frac{t_0 (1 + g^2) - g (\varepsilon_L + \varepsilon_R)}{1 - 2lg + g^2} \\ [3ex]
\displaystyle \tilde{k} = & \displaystyle \frac{k(1 + g^4) - 4gs(1 + g^2) + 4g^2j + 2g^2u}{(1 - 2lg + g^2)^2} \\ [3ex]
\displaystyle \tilde{j} = & \displaystyle \frac{j(1 + g^4) - 4gs(1 + g^2) + 2g^2(k + j + u)}{(1 - 2lg + g^2)^2} \\ [3ex]
\displaystyle \tilde{s} = & \displaystyle \frac{s(1 + 6g^2 + g^4) - g(1 + g^2)(2j + k + u)}{(1 - 2lg + g^2)^2}  \\ [3ex]
\displaystyle \tilde{u} = & \displaystyle \frac{u(1 + g^4) - 4gs(1 + g^2) + 4g^2j + 2g^2(k + 2j)}{(1 - 2lg + g^2)^2} .
\end{array}
\end{equation}

\section{Coulomb integrals}\label{app:Coulomb}

Most generally, we wish to evaluate the integral
\begin{equation}
I ({\bm P}, {\bm Q}) = \frac{1}{\pi^3a^4b^2}\int d^3r_1\int d^3r_2 \frac{e^{-\frac{(x_1 - P_x)^2 + (y_1 - P_y)^2}{a^2} - \frac{(z_1 -
P_z)^2}{b^2})} e^{-\frac{(x_2 - Q_x)^2 + (y_2 - Q_y)^2}{a^2} - \frac{(z_2 - Q_z)^2}{b^2})} }{|{\bm r}_1 - {\bm r}_2|}
\end{equation}
Define
\begin{equation}
\arraycolsep 0.3 ex
\begin{array}{rl}
\displaystyle f ({\bm r}) = & \displaystyle \int \frac{d^3k}{(2\pi)^3} \, e^{-i{\bm k}\cdot{\bm r}}F({\bm k}) \\ [3ex]
\displaystyle F ({\bm k}) = & \displaystyle \int d^3r \, e^{i{\bm k}\cdot{\bm r}}f({\bm r}).
\end{array}
\end{equation}
The Fourier transform of $f(\bm r) = 1/r$ is $F(\bm k) = 4\pi/k^2$. We write the term $1/|{\bm r}_1 - {\bm r}_2|$ as a Fourier expansion
\begin{equation}
\frac{1}{|{\bm r}_1 - {\bm r}_2|} = \frac{4\pi}{(2\pi)^3}\int d^3 k_2 \, \frac{e^{-i{\bm k}_2\cdot({\bm r}_1 - {\bm r}_2)}}{k_2^2} =
\frac{1}{2\pi^2}\int d^3 k_2 \, \frac{e^{-i{\bm k}_2\cdot({\bm r}_1 - {\bm r}_2)}}{k_2^2}.
\end{equation}
Next we Fourier transform all the other terms (Gaussians) entering $I ({\bm P}, {\bm Q})$. First we take the $z$-dependent terms
\begin{equation}
\arraycolsep 0.3 ex
\begin{array}{rl}
\displaystyle e^{-\frac{(z_1 - P_z)^2}{b^2}} = & \displaystyle \frac{b}{2\sqrt{\pi}}\int dk_{1z} \, e^{-ik_{1z}z_1}
e^{-\frac{b^2k_{1z}^2}{4}}e^{ik_{1z}P_z} \\ [3ex] \displaystyle e^{-\frac{(z_2 - Q_z)^2}{b^2}} = & \displaystyle \frac{b}{2\sqrt{\pi}}\int
dk_{3z} \, e^{-ik_{3z}z_2} e^{-\frac{b^2k_{3z}^2}{4}}e^{ik_{3z}Q_z}.
\end{array}
\end{equation}
We also Fourier transform the $x$ and $y$-dependent terms
\begin{equation}
\arraycolsep 0.3 ex
\begin{array}{rl}
\displaystyle e^{-\frac{({\bm r}_{1\perp} - {\bm P}_\perp)^2}{a^2}} = & \displaystyle \frac{a^2}{4\pi} \int \!\!\!\!\!  \int d^2k_{1\perp}
e^{-i{\bm k}_{1\perp}\cdot({\bm r}_{1\perp} - {\bm P}_\perp)} e^{- \frac{a^2k_{1\perp}^2}{4}} \\ [3ex]
\displaystyle e^{-\frac{({\bm r}_{2\perp}
- {\bm Q}_\perp)^2}{a^2}} = & \displaystyle \frac{a^2}{4\pi} \int \!\!\!\!\! \int d^2k_{3\perp} e^{-i{\bm k}_{3\perp}\cdot({\bm r}_{2\perp} -
{\bm Q}_\perp)} e^{- \frac{a^2k_{3\perp}^2}{4}}.
\end{array}
\end{equation}
The integral $I ({\bm P}, {\bm Q})$ becomes
\begin{equation}
\arraycolsep 0.3 ex
\begin{array}{rl}
\displaystyle I = & \displaystyle \int \!\!\!\!\! \int\!\!\!\!\! \int\!\!\!\!\! \int\!\!\!\!\! \int \frac{d^3r_1 d^3r_2d^3 k_1 d^3k_2
d^3k_3}{128\pi^8} \, \frac{e^{-i({\bm k}_1 + {\bm k}_2)\cdot{\bm r}_1}e^{-i ({\bm k}_3 - {\bm k}_2)\cdot{\bm r}_2}}{k_2^2} \, e^{i{\bm k}_1\cdot
{\bm P}} e^{i{\bm k}_3 \cdot {\bm Q}} e^{- \frac{(a^2k_{1\perp}^2 + b^2k_{1z}^2)}{4}} e^{- \frac{(a^2k_{3\perp}^2 + b^2k_{3z}^2)}{4}}
\\ [3ex]
= & \displaystyle \frac{1}{2\pi^2} \int \frac{d^3 k_1}{k_1^2} \, e^{i{\bm k}_1\cdot ({\bm P}
- {\bm Q})} e^{- \frac{(a^2k_{1\perp}^2 + b^2k_{1z}^2)}{2}}.
\end{array}
\end{equation}
The integral depends only on the difference ${\bm R} = {\bm Q} - {\bm P}$. If ${\bm R}$ is in the $xy$-plane we obtain
\begin{equation}
\arraycolsep 0.3 ex
\begin{array}{rl}
\displaystyle I ({\bm R}) = \int_0^\infty dk_\perp \, e^{- \frac{(a^2 - b^2)k_\perp^2}{2}} J_0 (k_\perp R) \, Erfc
\bigg(\frac{bk_\perp}{\sqrt{2}}\bigg).
\end{array}
\end{equation}
In 2D, i.e. as $b \rightarrow 0$, we have $I_{2D}(0) = \frac{1}{a}\sqrt{\frac{\pi}{2}}$.  In Hund-Mulliken problems we typically require
evaluation of four bare Coulomb integrals (direct and exchange, on-site repulsion, plus an additional term), namely
\begin{equation}
\arraycolsep 0.3 ex
\begin{array}{rl}
\displaystyle k = & \displaystyle \frac{e^2}{\epsilon} \, I (- X_0 \hat{\bm x}, X_0 \hat{\bm x}) \equiv \frac{e^2}{\epsilon}  \, I (2X_0
\hat{\bm x}) \\ [3ex] \displaystyle j = &
\displaystyle \frac{e^2}{\epsilon} \, l^2 \, I (X_0 \hat{\bm x}, X_0 \hat{\bm x}) \equiv
\frac{e^2}{\epsilon} \, l^2 \, I (0) \\ [3ex]
\displaystyle u = & \displaystyle \frac{e^2}{\epsilon} \, I (X_0 \hat{\bm x}, X_0 \hat{\bm x})
\equiv \frac{e^2}{\epsilon} \, I (0) \\ [3ex]
\displaystyle s = & \displaystyle \frac{e^2}{\epsilon} \, l \, I (- X_0 \hat{\bm x}, 0) \equiv
\frac{e^2}{\epsilon} \, l \, I (X_0 \hat{\bm x}).
\end{array}
\end{equation}

\section{Matrix elements in a magnetic field}
\label{sec:mag}

In a magnetic field the wave function of each dot acquires an additional phase, and we write them as $L_{\pm m}$, $R_{\pm m}$
\begin{equation}
\arraycolsep 0.3 ex
\begin{array}{rl}
\displaystyle L_{\pm m} = & \displaystyle e^{-iyX_0/2l_B^2} L_{\pm} \\ [3ex]
\displaystyle R_{\pm m} = & \displaystyle e^{iyX_0/2l_B^2} R_{\pm}.
\end{array}
\end{equation}
The magnetic length is $l_B = \sqrt{\hbar/eB}$, the Larmor frequency $\omega_L = eB/2m$, the effective frequency $\omega = \sqrt{\omega_0^2 +
\omega_L^2}$ and the parameter $b_m = \sqrt{1 + \omega_L^2/\omega_0^2}$, while $a$, $d$ and $l$ all change to $a_m = a/\sqrt{b_m}$, $d_m =
d\sqrt{b_m}$, and $l_m = e^{-d^2(2b_m - 1/b_m)}$. The Coulomb integral $I$  changes to $I_m$, in which all the parameters are replaced by their
corresponding expressions in a magnetic field, so $a \rightarrow a_m$ and so forth. The Coulomb matrix elements change as follows
\begin{equation}
\arraycolsep 0.3 ex
\begin{array}{rl}
\displaystyle k_m = & \displaystyle \frac{e^2}{\epsilon} \, I_m (2X_0 \hat{\bm x}) \\ [3ex]
\displaystyle j_m = & \displaystyle \frac{e^2}{\epsilon} \, l_m^2 \, I_m (- \frac{iX_0a^2_m}{l_B^2}\hat{\bm y}) \\ [3ex]
\displaystyle u_m = & \displaystyle \frac{e^2}{\epsilon} \, I_m (0) \\ [3ex]
\displaystyle s_m = & \displaystyle \frac{e^2}{\epsilon} \, l_m \, I_m (X_0 \hat{\bm x} - \frac{iX_0a^2_m}{2l_B^2}\hat{\bm y}) ).
\end{array}
\end{equation}
The bare tunneling parameter $t_0$ changes to $e^{-\frac{X_0^2a_m^2}{4l_B^4}}t_m$, with $t_m$ found from $t_0$ by replacing $a$ with $a_m$ and
so forth.

\section{Valley-exchange Coulomb integral}
\label{sec:valleyx}

The valley-exchange Coulomb integral for a dot located at $ {\bm R}_D = 0$
\begin{equation}
j_v = \int d^3 r_1 \int d^3 r_2\, D_z^{*(1)} D_{\bar{z}}^{*(2)} V_{ee} \, D_{\bar{z}}^{(1)} D_z^{(2)}.
\end{equation}
The aim is to reduce this integral to the form of \textit{expression} $\times I({\bm P}, {\bm Q})$ as calculated above. Firstly,
\begin{equation}
D_z^{*(1)} D_{\bar{z}}^{*(2)} V_{ee} \, D_{\bar{z}}^{(1)} D_z^{(2)} = \frac{1}{\pi^{3} a^4b^2}\, \frac{e^2}{\epsilon|{\bm r}_1 - {\bm r}_2|} \,
e^{-\frac{x_1^2 + y_1^2}{a^2} -\frac{z_1^2}{b^2}}  e^{ - 2i{\bm k}_{z} \cdot ({\bm r}_1 - {\bm r}_2)} e^{-\frac{x_2^2 + y_2^2}{a^2}
-\frac{z_2^2}{b^2}} u_z^* ({\bm r}_1) u_{\bar{z}} ({\bm r}_1) u_{\bar{z}}^* ({\bm r}_2) u_z({\bm r}_2).
\end{equation}
We expand the lattice-periodic wave functions as
\begin{equation}\nonumber
\arraycolsep 0.3 ex
\begin{array}{rl}
\displaystyle u_{z}^*({\bm r}_1) = & \displaystyle \sum_{{\bm K}_1} c^{z*}_{{\bm K}_1} e^{-i{\bm K}_1 \cdot{\bm r}_1} \\ [3ex]

\displaystyle u_{\bar{z}}({\bm r}_1) = & \displaystyle \sum_{{\bm K}_3} c^{\bar{z}}_{{\bm K}_3} e^{i{\bm K}_3 \cdot{\bm r}_1} \\ [3ex]

\displaystyle u_{\bar{z}}^*({\bm r}_2) = & \displaystyle \sum_{{\bm K}_2} c^{\bar{z}*}_{{\bm K}_2} e^{-i{\bm K}_2 \cdot{\bm r}_2} \\ [3ex]

\displaystyle u_{z}({\bm r}_2) = & \displaystyle \sum_{{\bm K}_4} c^z_{{\bm K}_4} e^{i{\bm K}_4 \cdot{\bm r}_2} \\ [3ex]

\displaystyle u_z^* ({\bm r}_1) u_{\bar{z}} ({\bm r}_1) u_{\bar{z}}^* ({\bm r}_2) u_z({\bm r}_2) = & \displaystyle \sum_{{\bm K}_1, {\bm K}_2,
{\bm K}_3, {\bm K}_4} c^{z*}_{{\bm K}_1} c^{\bar{z}}_{{\bm K}_3}  c^{\bar{z}*}_{{\bm K}_2} c^z_{{\bm K}_4} e^{-i({\bm K}_1 - {\bm K}_3)
\cdot{\bm r}_1} e^{-i ({\bm K}_2 - {\bm K}_4)\cdot{\bm r}_2}
\\ [3ex]
= & \displaystyle \sum_{{\bm K}_1, {\bm K}_2, {\bm \kappa}_1, {\bm \kappa}_2} c^{z*}_{{\bm K}_1} c^{\bar{z}}_{{\bm K}_1 - {\bm \kappa}_1}
c^{\bar{z}*}_{{\bm K}_2} c^z_{{\bm K}_2 - {\bm \kappa}_2} e^{-i{\bm \kappa}_1 \cdot{\bm r}_1} e^{-i {\bm \kappa}_2  \cdot{\bm r}_2},
\end{array}
\end{equation}
where ${\bm \kappa}_1 = {\bm K}_1 - {\bm K}_3$ and ${\bm \kappa}_2 = {\bm K}_2 - {\bm K}_4$. Abbreviating $\sum \equiv \sum_{{\bm K}_1, {\bm
K}_2, {\bm \kappa}_1, {\bm \kappa}_2} $, ${\bm \kappa}_1 + 2{\bm k}_{z} = \bar{{\bm \kappa}}_1$ and $\bar{{\bm \kappa}}_2 = {\bm \kappa}_2 -
2{\bm k}_{z}$
\begin{equation}
D_z^{*(1)} D_{\bar{z}}^{*(2)} V_{ee} \, D_{\bar{z}}^{(1)} D_z^{(2)} = \frac{\sum c^{z*}_{{\bm K}_1} c^{\bar{z}*}_{{\bm K}_2} c^{\bar{z}}_{{\bm
K}_1 - {\bm \kappa}_1 } c^z_{{\bm K}_2 - {\bm \kappa}_2}}{\pi^{3} a^4b^2}\, \bigg(e^{-\frac{x_1^2 + y_1^2}{a^2} - \frac{z_1^2}{b^2}}  e^{ - i
\bar{{\bm \kappa}}_1 \cdot {\bm r}_1} \, \frac{e^2}{\epsilon|{\bm r}_1 - {\bm r}_2|} \, e^{-\frac{x_2^2 + y_2^2}{a^2} -\frac{z_2^2}{b^2}} e^{ -i
\bar{{\bm \kappa}}_2 \cdot {\bm r}_2}\bigg).
\end{equation}
The integrand is almost in the form we want it, except we need to complete the square in the Gaussians.  To shorten the algebra a little we note
that the only sizable terms in the sum over reciprocal lattice wave vectors will have $\bar{\kappa}_{1x} = \bar{\kappa}_{1y} = \bar{\kappa}_{2x}
= \bar{\kappa}_{2y} = 0$. All the rest of the terms are exponentially suppressed. Consequently we only need
\begin{equation}
\arraycolsep 0.3 ex
\begin{array}{rl}
\displaystyle \frac{1}{b^2}(z_{1, 2}^2 + ib^2\bar{\kappa}_{1z, 2z}) = &  \displaystyle \frac{1}{b^2}\bigg[\bigg(z_{1,2} +
\frac{ib^2\bar{\kappa}_{1z, 2z}}{2}\bigg)^2 + \frac{b^4\bar{\kappa}_{1z, 2z}^2}{4}\bigg].
\end{array}
\end{equation}
Therefore the valley exchange Coulomb integral can be expressed as
\begin{equation}
j_v = \frac{e^2}{\epsilon}\sum_{{\bm K}_1, {\bm K}_2, {\bm \kappa}_1, {\bm \kappa}_2} c^{z*}_{{\bm K}_1} c^{\bar{z}*}_{{\bm K}_2}
c^{\bar{z}}_{{\bm K}_1 - {\bm \kappa}_1} c^z_{{\bm K}_2 - {\bm \kappa}_2 }\, e^{-\frac{b^2(\bar{\kappa}_{1z}^2 + \bar{\kappa}_{2z}^2)}{4}} \, I
\bigg[\frac{-ib^2(\bar{\kappa}_{1z} - \bar{\kappa}_{2z})}{2} \, \hat{\bm z} \bigg].
\end{equation}
Using the result for the integral $I$ found above and $\Delta \bar{\kappa}_z = \bar{\kappa}_{1z} - \bar{\kappa}_{2z}$
\begin{equation}
\arraycolsep 0.3 ex
\begin{array}{rl}
\displaystyle I \bigg[-\frac{ib^2\Delta\bar{\kappa}_z}{2} \, \hat{\bm z} \bigg] = & \displaystyle \frac{1}{\pi}\int_0^\infty dk_\perp \, k_\perp
\, e^{- \frac{a^2k_\perp^2}{2}} \int_{-\infty}^{\infty} dk_z\, \frac{e^{\frac{k_zb^2 \Delta\bar{\kappa}_z}{2}} e^{- \frac{b^2k_z^2}{2}}
}{k_\perp^2 + k_z^2} \\ [3ex] k_z^2 - k_z \Delta\bar{\kappa}_z = & \displaystyle \bigg(k_z - \frac{\Delta \bar{\kappa}_z}{2}\bigg)^2 -
\frac{\Delta \bar{\kappa}_z^2}{4}.
\end{array}
\end{equation}
We redefine $k_z \rightarrow k_z + \frac{\Delta\bar{\kappa}_z}{2}$ in the above integral without affecting the limits of integration
\begin{equation}
I \bigg[-\frac{ib^2\Delta\bar{\kappa}_z}{2} \, \hat{\bm z} \bigg] = \frac{e^{\frac{b^2 \Delta\bar{\kappa}_z^2}{8}} }{\pi}\int_0^\infty dk_\perp
\, k_\perp \, e^{- \frac{a^2k_\perp^2}{2}} \int_{-\infty}^{\infty} dk_z\, \frac{e^{- \frac{b^2k_z^2}{2}} }{k_\perp^2 + (k_z +
\frac{\Delta\bar{\kappa}_z}{2})^2}
\end{equation}
where $a k_\perp = q_\perp$, $b k_z = q_z$ and $b \Delta\bar{\kappa}_z = \Delta q_z$, so that
\begin{equation}
I \bigg[-\frac{ib^2\Delta\bar{\kappa}_z}{2} \, \hat{\bm z} \bigg] = \frac{e^{\frac{\Delta q_z^2}{8}} b}{\pi a^2}\int_0^\infty dq_\perp \,
q_\perp \, e^{- \frac{q_\perp^2}{2}} \int_{-\infty}^{\infty} dq_z\, \frac{e^{- \frac{q_z^2}{2}} }{q_\perp^2 \big(\frac{b^2}{a^2}\big) + (q_z +
\frac{\Delta q_z}{2})^2}.
\end{equation}
The exponent in the overall prefactor reduces to $- \frac{b^2 (\kappa_{1z} + \kappa_{2z})^2}{8}$, and the prefactor has no dependence on $k_0$.
The coefficients $c^\xi_{\bm K}$ for Si were calculated in Ref.~\onlinecite{Koiller_PRB04} using a pseudopotential method. For the final step in
the evaluation of the integral we examine which reciprocal lattice vectors make a sizable contribution. We need $\kappa_{1z} + \kappa_{2z} = 0$.
The first possibility is $\kappa_{1z} = \kappa_{2z} = 0$, with $\Delta q_z = 4bk_0$, which yields, using the coefficients found in Ref.\
\onlinecite{Koiller_PRB04}
\begin{equation}
\sum \approx \sum_{{\bm K}_1, {\bm K}_2} c^{z*}_{{\bm K}_1} c^{\bar{z}}_{{\bm K}_1}  c^{\bar{z}*}_{{\bm K}_2} c^z_{{\bm K}_2} \approx 0.01.
\end{equation}
The second possibility is $\kappa_{1z} = - \kappa_{2z}$, in which case $\Delta q_z  = 2b\kappa_{1z} + 4bk_0$.  These terms cancel exactly
because of lattice symmetry. For example, consider the case $\kappa_{1z} = 2 \, (2\pi/a_{Si})$ (that is, the Umklapp terms), for which
\begin{equation}
\sum_{Umklapp} \approx \sum_{{\bm K}_1, {\bm K}_2} c^{z*}_{{\bm K}_1} c^{\bar{z}}_{{\bm K}_1 - 2\, \frac{2\pi}{a_{Si}}\hat{\bm z}}
c^{\bar{z}*}_{{\bm K}_2} c^z_{{\bm K}_2 + 2\, \frac{2\pi}{a_{Si}}\hat{\bm z}}.
\end{equation}
Using the fact that $c_{-K}^{-z} = c_{K}^{z*}$, the sum of the largest coefficients entering $\sum_{Umklapp}$ is zero.
%












%
The valley-exchange Coulomb integral is approximately
\begin{equation}
j_v \approx \frac{e^2}{\epsilon a}\, \big(\sum\big)\, \bigg\{ \frac{b}{\pi a}\int_0^\infty dq_\perp \, q_\perp \, e^{- \frac{q_\perp^2}{2}}
\int_{-\infty}^{\infty} dq_z\, \frac{e^{- \frac{q_z^2}{2}} }{q_\perp^2 \big(\frac{b^2}{a^2}\big) + (q_z + \frac{\Delta q_z}{2})^2}\bigg\}
\end{equation}
For a = 8.2 nm, b = 3nm and $\epsilon \approx 7.9$, we get $\Delta q_z  \approx 117 $ and $(e^2/\epsilon a) \approx 22.2$meV.  The integral in
brackets is evaluated using Mathematica and yields $\approx 8.5\times 10^{-5}$, so the final result is $j_v \approx 0.02\mu$eV.

\end{widetext}


\end{document}